\documentclass[10pt]{article}
\pdfoutput=1
\usepackage[utf8]{inputenc}
\usepackage{cite}
\usepackage{amsmath,amssymb,amsbsy,amstext,amsthm,simplewick,amsfonts}
\usepackage{graphicx}
\usepackage{wrapfig}
\usepackage{upgreek}
\usepackage{bm} 
\usepackage{framed}
\usepackage{bbm}
\usepackage{textcomp}
\usepackage{tikz}
\usepackage{pifont}
\usetikzlibrary{matrix,shapes,fit,tikzmark,calc}
\usepackage{adjustbox}
\usepackage{makecell}
\usepackage{tcolorbox}
\usepackage{physics}
\usepackage{empheq}
\usepackage[normalem]{ulem}
\usepackage{enumitem}
\usepackage{braket} 
\usepackage{array}
\usepackage{dsfont}
\usepackage{ulem}

\numberwithin{equation}{section}

\def\be{\begin{equation}}
\def\ee{\end{equation}}

\def\ba{\begin{eqnarray}}
\def\ea{\end{eqnarray}}

\def\O{\Omega}
\def\bfx{\textbf{x}}

\def\bfk{\textbf{k}}
\def\bfx{\textbf{x}}
\def\bfq{\textbf{q}}

\usepackage{caption}
\usepackage{subcaption}

\usepackage[framemethod=default]{mdframed}
\newmdenv[skipabove=7pt,
skipbelow=7pt,
rightline=false,
leftline=false,
topline=false,
bottomline=false,
backgroundcolor=gray!10,
linecolor=gray,
innerleftmargin=5pt,
innerrightmargin=5pt,
innertopmargin=5pt,
innerbottommargin=5pt,
leftmargin=0cm,
rightmargin=0cm,
linewidth=4pt]{eBox}
\newmdenv[skipabove=7pt,
skipbelow=7pt,
rightline=false,
leftline=false,
topline=false,
bottomline=false,
backgroundcolor=gray!10,
linecolor=gray,
innerleftmargin=5pt,
innerrightmargin=5pt,
innertopmargin=-5pt,
innerbottommargin=5pt,
leftmargin=0cm,
rightmargin=0cm,
linewidth=4pt]{eBox2}

\definecolor{blue3}{RGB}{31,119,180}
\definecolor{red3}{RGB}{214,39,40}
\definecolor{orange3}{RGB}{255,127,14}
\definecolor{green3}{RGB}{44,160,44}

\usepackage{colortbl}
\definecolor{lightgreen}{cmyk}{0.2, 0, 0.2, 0.2}
\definecolor{lightgray}{cmyk}{0.1,0.2,0,0.1}
\definecolor{lightgray2}{cmyk}{0.1,0.1,0,0.1}

\usepackage{hyperref}
\hypersetup{colorlinks=true,linkcolor=teal,citecolor=orange3,urlcolor=green3,pdfencoding=auto}

\makeatletter
\newlength{\apb@width}
\newcommand{\autoparbox}[2][c]{\settowidth{\apb@width}{#2}\parbox[#1]{\apb@width}{#2}}

\makeatother


\def\O{{\cal O}}

\def\nn{\nonumber}
\def\eps{\boldsymbol\epsilon}

\def\beq{\begin{equation}}
\def\eeq{\end{equation}}

\allowdisplaybreaks[1]
\begin{document}


\begin{titlepage}
\setcounter{page}{1} \baselineskip=15.5pt 
\thispagestyle{empty}

\begin{center}
{\fontsize
{28}{28}\centering \bf On Graviton non-Gaussianities in the \\ \vspace{0.5cm} Effective Field Theory of Inflation \;}\\
\end{center}

\vskip 18pt
\begin{center}
\noindent
{\fontsize{12}{18}\selectfont Giovanni Cabass\footnote{\tt gcabass@ias.edu}$^{,a}$, David Stefanyszyn\footnote{\tt david.stefanyszyn@nottingham.ac.uk}$^{,b,c}$, Jakub Supe{\l}\footnote{\tt js2154@cam.ac.uk}$^{,d}$ and Ayngaran Thavanesan\footnote{\tt at735@cam.ac.uk}$^{,e}$}
\end{center}

\begin{center}
\vskip 8pt
$a$\textit{ School of Natural Sciences, Institute for Advanced Study, Princeton, NJ 08540, United States} \\ 
$b$ \textit{School of Mathematical Sciences \& School of Physics and Astronomy,
University of Nottingham, University Park, Nottingham, NG7 2RD, UK} \\
$c$ \textit{Nottingham Centre of Gravity, University of Nottingham, University Park, Nottingham, NG7 2RD, UK} \\
$d$\textit{ Department of Applied Mathematics and Theoretical Physics, University of Cambridge, Wilberforce Road, Cambridge, CB3 0WA, UK} \\
$e$ \textit{Kavli Institute for Cosmology, Madingley Road, Cambridge, CB3 0HA, UK}
\end{center}


\vspace{1.4cm}

\noindent We derive parity-even graviton bispectra in the Effective Field Theory of Inflation (EFToI) to all orders in derivatives. Working in perturbation theory, we construct all cubic interactions that can contribute to tree-level graviton bispectra, showing that they all come from EFToI operators containing two or three powers of the extrinsic curvature and its covariant derivatives: all other operators can be removed by field redefinitions or start at higher-order in perturbations. For operators cubic in the extrinsic curvature, where the single-clock consistency relations are satisfied without a correction to the graviton two-point function, we use the Manifestly Local Test (MLT) to efficiently extract the effects of evolving graviton fluctuations to the end of inflation. Despite the somewhat complicated nature of the bulk interactions, the final boundary correlators take a very compact form. For operators quadratic in the extrinsic curvature, the leading order bispectra are a sum of contact and single exchange diagrams, which are tied together by spatial diffeomorphisms, and to all orders in derivatives we derive these bispectra by computing the necessary bulk time integrals. For single exchange diagrams we exploit factorisation properties of the bulk-bulk propagator for massless gravitons and write the result as a finite sum over residues. Perhaps surprisingly, we show these single exchange contributions have only total-energy poles and also satisfy the MLT.


\end{titlepage} 


\newpage
\setcounter{tocdepth}{2}
{
\hypersetup{linkcolor=black}
\tableofcontents
}

\newpage


\section{Introduction}

\noindent Graviton self-interactions are heavily constrained in exact de Sitter space \cite{Maldacena:2011nz}. The power of full de Sitter isometries ensures that there are only three contributions to the cubic part of the Wavefunction of the Universe (WFU): two of these are parity-even and arise from the Einstein-Hilbert action and a six-derivative correction in the form of a Riemann cubed operator, while the other is parity-odd and comes from a parity-odd Riemann cubed operator. In perturbation theory, expectation values of graviton fields at the late-time boundary of de Sitter space can be extracted from knowledge of the \textit{wavefunction coefficients} that appear in the WFU and upon computing the corresponding bispectra, only the two parity-even contributions survive \cite{Soda:2011am,Shiraishi:2011st}\footnote{The parity-odd interaction contributes only a pure phase to the wavefunction so drops out when we compute expectation values.}. This rigidity of graviton interactions in de Sitter space very nicely parallels a similar story for the $S$-matrix in Minkowski space \cite{Benincasa:2007qj,TASI,Benincasa:2007xk,PSS}. \\

\noindent During inflation, however, additional \textit{shapes} of graviton bispectra can arise due to the breaking of de Sitter boosts. Indeed, in the Effective Field Theory of Inflation (EFToI) \cite{Cheung:2007st} a scalar field with a time-dependent background profile spontaneously breaks de Sitter boosts thereby allowing for a richer structure of graviton self-interactions which only need to be invariant under spatial diffeomorphisms. The graviton action up to cubic order in the transverse, traceless fluctuation $\gamma_{ij}$ is fixed to be that of the Einstein-Hilbert action to leading order in derivatives with corrections containing at least three derivatives \cite{Creminelli:2014wna,Bordin:2017hal}. The leading corrections to the Einstein-Hilbert action, and the resulting bispectra, have been computed in \cite{CabassBordin,Cabass:2021fnw} where there are two new shapes: one comes from a parity-even operator while the second comes from a parity-odd operator. Some four-derivative operators have also been explored in \cite{CabassBordin}. One can go even further and break spatial diffeomorphisms as in e.g. Solid Inflation \cite{SolidInflation} to yield even more self-interactions, but now in the presence of a (small) graviton mass. A \textit{zoology} of graviton interactions during inflation has been derived in \cite{ZoologyNG} where it is clear that even a small breaking of maximal symmetries opens up many new possibilities.  \\

\noindent In this work we aim to perform a more general analysis within the EFToI. Working in perturbation theory, and assuming invariance under spatial translations, rotations and scale invariance, we construct parity-even graviton bispectra to all orders in derivatives. Throughout we work within the WFU formalism and derive expectation values only at the end of our computations, and as a consistency check we verify that our results satisfy the leading order consistency relations of single-clock cosmologies \cite{Hinterbichler:2013dpa,Maldacena:2002vr,Creminelli:2012ed,Hinterbichler:2012nm}. The highly constrained parity-odd bispectra in the EFToI have been computed in \cite{CabassBordin,Cabass:2021fnw,Bartolo:2017szm,Bartolo:2020gsh}, and in this work we fill the parity-even gap. For parity-even interactions the number of bispectra grows unbounded as we increase the number of derivatives, which is in stark contrast to the parity-odd situation where one can use unitarity methods to prove that only a single tree-level shape for each helicity configuration is allowed \cite{Cabass:2021fnw}\footnote{Even if we further break spatial diffeomorphism as in e.g. Solid inflation, only three shapes are allowed \cite{Cabass:2021fnw} (see also \cite{Orlando:2022rih}).}. \\

\noindent Let us explain our motivation for this work. Given that gravitational interactions are so heavily constrained in exact de Sitter space, it is interesting to carve out the space of consistent self-interactions during inflation where the most powerful de Sitter symmetries are broken. The EFToI is the natural place to start given that the unbroken spatial diffeomorphisms are still expected to provide non-trivial relations between different operators, and they ensure that the graviton remains massless unlike in Solid Inflation \cite{SolidInflation} and massive gravity \cite{WFCtoCorrelators2}. Our long-term objective is to gain a solid understanding of Quantum Field Theory on inflationary backgrounds and graviton $n$-point functions are certainly objects of interest. In flat-space, the tree-level four-point scattering amplitude for massless gravitons at low energies is completely fixed by symmetries, locality and consistent factorisation (see e.g. \cite{Cheung:2017pzi,Benincasa:2007xk}) while higher-point amplitudes can be extracted from three-point ones using BCFW momentum shifts \cite{BCFW,Benincasa:2007qj}. Boost-breaking amplitudes have also been constructed in \cite{PSS,DSJS}. Striving for a comparable understanding of cosmological $n$-point functions is one of the primary goals of the bootstrap approach to constructing cosmological observables \cite{Maldacena:2011nz,Arkani-Hamed:2015bza,COT,Melville:2021lst,Goodhew:2021oqg,Cespedes:2020xqq,MLT,Baumann:2021fxj,Meltzer:2021zin,Gomez:2021qfd,Sleight:2021plv,DiPietro:2021sjt,Hogervorst:2021uvp,CosmoBootstrap1,CosmoBootstrap2,CosmoBootstrap3,Bonifacio:2021azc,PSS,BBBB,Benincasa:2019vqr,Benincasa:2018ssx,Arkani-Hamed:2017fdk,Green:2020ebl,Sleight:2021iix,Sleight:2020obc,Sleight:2019hfp,Sleight:2019mgd,Hillman:2019wgh,Bittermann:2022nfh,Benincasa:2021qcb,Baumann:2022jpr,Benincasa:2022gtd,Benincasa:2022omn,Armstrong:2022csc,Pimentel:2022fsc,Jazayeri:2022kjy,Bzowski:2012ih,Bzowski:2011ab,Bzowski:2013sza,Mata:2012bx,Kundu:2014gxa,Kundu:2015xta,Ghosh:2014kba,Heckelbacher:2022hbq,Benincasa:2020aoj,Qin:2022fbv}, and this endeavour requires theoretical data from which we can learn about the allowed structures. Ultimately we would like to distinguish between different inflationary models directly at the level of observables and we hope that our work in this paper will contribute to this goal.  \\

\noindent In some sense our work builds on \cite{Cabass:2021fnw} where graviton bispectra were constructed from very general principles. Using the aforementioned symmetries (no de Sitter boosts), locality and unitarity, all possible shapes were derived under the assumption that the graviton mode functions are the usual massless de Sitter ones. Symmetries and the assumption of a Bunch-Davies vacuum\footnote{For a discussion on graviton bispectra with non-Bunch-Davies initial states see e.g. \cite{nonBD}.} were used to write down a general ansatz for tree-level wavefunction coefficients which can then be constrained by the Manifestly Local Test (MLT) derived in \cite{MLT}. This test demands that wavefunction coefficients, when expanded for small energies\footnote{Even though in cosmology we don't have time translation symmetry, we follow standard convention and still refer to the norms of spatial momenta as energies.}, must not contain a term linear in any of the energies and follows from a simple property of the massless mode functions. Unitarity, in the form of the Cosmological Optical Theorem (COT) \cite{COT}, was used to deduce which part of the wavefunction coefficients contribute to expectation values and in particular to the bispectra. For parity-even interactions, both rational and logarithmic contributions to the wavefunction coefficients also contribute to the bispectra, while for parity-odd interactions only the coefficients of logarithms contribute. This makes parity-odd bispectra highly-constrained, since logarithms in the wavefunction are rare, and indeed only three shapes are allowed with only one linear combination appearing in the EFToI. In \cite{Cabass:2021fnw} no assumption was made about how de Sitter boosts are spontaneously broken, i.e. no assumption was made about the details of the underlying inflationary model, and our aim in this paper is to derive the parity-even EFToI subset of these general graviton bispectra. \\

\noindent To achieve this goal we use a combination of bulk and bootstrap tools. We start with the general action of the EFToI \cite{Cheung:2007st} and show that all on-shell graviton three-point functions can be derived from corrections to the Einstein-Hilbert action that are built out of the extrinsic curvature and its covariant derivatives only. We use geometric identities and field redefinitions to arrive at this conclusion. Operators that can contribute to graviton bispectra at tree-level are at most cubic in the extrinsic curvature, and we derive all quadratic and cubic vertices for the transverse, traceless graviton that come from these covariant operators. Operators that are quadratic in the extrinsic curvature correct both the quadratic and cubic action, while operators that are cubic in the extrinsic curvature correct only the cubic vertices. In this latter case the only contributions to the bispectra at tree-level come from WFU contact diagrams and to compute these diagrams we use the techniques of \cite{Cabass:2021fnw}. We are able to write down closed form expressions to all orders in derivatives with the freedom reduced to symmetric polynomials in the external energies with their degrees fixed by scale invariance. We refer to these bispectra as Type-II. In the former case both contact and single exchange diagrams can contribute and for reasons that will be explained in detail in the main body of this paper, we explicitly compute the bulk time integrals to extract the bispectra in this case. For exchange diagrams we use the representation of \cite{Meltzer:2021zin}, which relies on a neat factorisation property of the bulk-bulk propagator, and present the results in a very compact form to all orders in derivatives. We refer to these bispectra as Type-I. Such single exchange diagrams have also been very recently studied and bootstrapped in the context of the cosmological collider in \cite{Pimentel:2022fsc,Jazayeri:2022kjy}. For contact and exchange diagrams alike, the largest contribution in perturbation theory comes from diagrams that are linear in the couplings so we focus only on these. As pointed out in \cite{Cabass:2021fnw}, Type-II bispectra can be large thanks to the weaker constraints on their size coming from the validity of perturbation theory since they do not come with a correction to the graviton two-point function.  \\

\noindent Along the way we show that single exchange diagrams that contribute to the Type-I bispectra are only singular when the total energy goes to zero. This is somewhat surprising since in general such a diagram could have two singular points arising when the sum of energies entering either of the two vertices goes to zero. However, the quadratic corrections in the EFToI are such that the additional singularities are always cancelled. In hindsight this might have been expected since any other singularities would seem to violate the EFToI consistency relations, and one could also apply the locality arguments of \cite{BBBB} to arrive at the same conclusion. We also show that these single exchange diagrams satisfy the MLT. This is certainly not true for general exchange diagrams that arise due to quadratic mixing, see e.g. \cite{Pimentel:2022fsc,Jazayeri:2022kjy}, but the vertices in the EFToI ensure that the MLT is always satisfied. These two facts ensure that all the crucial assumptions of the analysis in \cite{Cabass:2021fnw} remain true for both contact and exchange diagrams, so both types of bispectra that we will derive in this paper are captured by that analysis. The easiest way to directly distinguish between Type-I and Type-II bispectra is to take the soft limit and check the consistency relations \cite{Hinterbichler:2013dpa,Maldacena:2002vr,Creminelli:2012ed,Hinterbichler:2012nm}, since for Type-I the power spectrum needs to be modified to satisfy them, while Type-II bispectra satisfy the consistency relations without the need for such a correction.\\

\noindent The rest of the paper is organised as follows. In the following section, Section \ref{EFToI}, we very briefly review the EFToI and show, for the first time, that all graviton bispectra can be derived from extrinsic curvature operators only. In Section \ref{GravitonInteractions} we derive the general Lagrangian for the transverse, traceless graviton that comes from these operators and can contribute to wavefunction coefficients and bispectra. In Section \ref{Generalities} we comment further on the Feynman diagrams that contribute to the cubic part of the wavefunction and show that the exchange diagrams have the same singularity structure as the contact ones, and that they satisfy the MLT. In this section we also remind the reader how cosmological correlators are extracted from wavefunction coefficients. In Sections \ref{TypeI} and \ref{TypeII} respectively, we construct the Type-I and Type-II bispectra, present compact expressions to all orders in derivatives, and verify that the leading order consistency relations are satisfied. Finally, we conclude in Section \ref{Conclusions}. 

\paragraph{Notation and conventions} \label{Conventions}
Throughout we work with the mostly positive metric signature and our Fourier transformation is defined as
\begin{align}
f(\bfx)&=\int \dfrac{d^3\bfk}{(2\pi)^3}{f}(\bfk)\exp(i\bfk\cdot\bfx)\equiv\int_{\bfk}{f}(\bfk)\exp(i\bfk\cdot\bfx) \label{FT} \,,\\
{f}(\bfk)&=\int d^3\bfx \,f(\bfx)\exp(-i\bfk\cdot\bfx)\equiv \int_{\bfx}  f(\bfx)\exp(-i\bfk\cdot\bfx)\,.
\end{align}
We parameterise the Wavefunction of the Universe $\Psi$ at conformal time $\eta_0$ in terms of the graviton's helicities as
\begin{align}\label{WavefunctionOfTheUniverse}
\Psi[\eta_{0},\gamma({\bf k})] = \text{exp}\left[-\sum_{n=2}^{\infty} \frac{1}{n!} \sum_{h_i=\pm} \int_{{\bfk}_{1}, \ldots, {\bf{k}}_{n}} \psi^{h_{1} \ldots h_{n}}_{n}({\bf k_{1}} \ldots {\bf k_{n}}) (2\pi)^3 \delta^3 \left(\sum \bfk_a \right) \gamma^{h_1}({\bf k}_{1}) \ldots \gamma^{h_n}({\bf k}_{n}) \right].
\end{align}
Here the ever-present momentum conserving delta function is a consequence of the unbroken spatial translations and we write this explicitly so it is not included in the wavefunction coefficients $\psi_{n}$. We write the corresponding cosmological correlators as $B_{n}$ (also with the delta function dropped). The relations between $B_{n}$ and $\psi_{n}$ will be given in the main body of the paper. In this WFU expression, $\gamma^h(\bfk)$ is the spin-$h$ Fourier mode of the graviton and is related to the position space graviton by
\begin{align}\label{gammaFT}
\gamma_{ij}(x,\eta)=\int_{\bfk} e^{i \bfk \cdot \bfx} \sum_{h=\pm} e_{ij}^{h}(\bfk) \gamma_{h}(\bfk,\eta) \,,
\end{align}
and for us $\gamma_{h}(\bfk,\eta)$ is given by the usual de Sitter mode functions. The graviton polarisation tensor satisfies the following relations:
\begin{align}\label{pol1}
e_{ii}^{h}(\bfk)&=k^{i} e^{h}_{ij}(\bfk)=0 & \text{(transverse and traceless)}\,,\\
e_{ij}^{h}(\bfk)&= e_{ji}^{h}(\bfk)& \text{(symmetric)} \,, \\
e_{ij}^{h}(\bfk) e_{jk}^{h}(\bfk)&=0 & \text{(lightlike)} \,, \\
e^{h}_{ij}(\bfk) e^{h'}_{ij}(\bfk)^{\ast}&=4\delta_{hh'} & \text{(normalization)}\,, \label{polnorm}  \\
e_{ij}^h(\bfk)^{\ast}&= e_{ij}^h(-\bfk) &\text{($  \gamma_{ij}(x) $ is real)}  \,.\label{poln}
\end{align} 
We will often encounter polynomials in the three external energies that are fully symmetric and we will express these in terms of the elementary symmetric polynomials:
\begin{align} 
k_{T} &= k_{1}+k_{2}+k_{3}, \\
e_{2} &= k_{1}k_{2}+k_{1}k_{3}+k_{2}k_{3}, \\
e_{3} &= k_{1}k_{2}k_{3}. \label{SymmetricPolys}
\end{align}


\section{Effective Field Theory of Inflation} \label{EFToI}

We work within the EFToI where the inflationary background is driven by a single scalar degree of freedom, the inflaton. In the language of symmetry breaking, this set up corresponds to the case where only time diffeomorphisms are broken by the background i.e. we couple a superfluid to gravity \cite{ZoologyNG}. Time diffeomorphisms are nevertheless non-linearly realised by the inflationary perturbations. The EFToI provides a formalism to capture the most general action for the graviton and scalar fluctuations on a quasi-de Sitter background, with operators organised in a derivative expansion and tadpole cancellation guaranteed to all orders. We begin this section by briefly reviewing the EFToI before showing that graviton vertices up to cubic order are captured by operators built out of the extrinsic curvature only. \\

\noindent Since we are interested in graviton interactions, we work with the following line element
\begin{equation} \label{FRWMetric}
ds^2 = -dt^2 + a^2(t) (e^{\gamma})_{ij} dx^{i}dx^{j},
\end{equation}
where $a = a(t)$ is the scale factor and we define the Hubble parameter $H = \dot{a}/ a$. For now we work in cosmological time but later on we will convert to conformal time which is more suitable for computing late-time cosmological correlators. Relative to the usual ADM formalism, we have set the lapse variable to unity and the shift together with the curvature perturbation $\zeta$ to zero. Usually we would integrate out the non-dynamical parts of the metric which would introduce additional interactions for $\gamma$ and $\zeta$, but up to cubic order this procedure does not alter the $\gamma$ interactions so we can safely work with \eqref{FRWMetric}\footnote{Since the graviton does not mix with the non-dynamical modes at linear order, it must appear at least quadratically in the solutions to the constraint equations. Plugging these solutions back into the action can therefore only affect the graviton interactions at quartic order and higher.}. We fix the gauge completely by taking the graviton to be transverse and traceless: $\gamma_{ii} = 0 = \partial_{i} \gamma_{ij}$\footnote{Note that in this gauge the metric determinant is independent of $\gamma_{ij}$ and is fixed by its background value.}. This gauge is usually referred to as unitary gauge where all degrees of freedom live in the metric i.e. the inflaton perturbation has been eaten by the metric. In this gauge the most general action that we can write down is the one consistent with spatial diffeomorphisms \cite{Cheung:2007st}:
\begin{eqnarray}
S = \int d^4 x \sqrt{-g}\, F(R_{\mu\nu\rho\sigma}, g^{00}, K_{\mu\nu}, \nabla_{\mu}, t),
\end{eqnarray}
where all free indices are, in general, upper $0$'s. For our purposes, however, there are simplifications. All $t$ dependence will be fixed by scale invariance, since we are working in the scale invariant approximation, i.e. we assume that $H$ and $\dot{H}$ vary slowly and restrict all $t$ dependence to be that coming from the metric. In practice this means that we work with a fixed de Sitter background metric, so all correlators we compute are valid up to small slow-roll corrections \cite{Maldacena:2002vr}. Scale invariance will be most transparent when we convert to conformal time. This is a technically natural set-up since in the $\phi$ language it corresponds to an approximate shift symmetry for the inflaton. We also take $g^{00} = -1$, since we don't include the lapse, and this means we can write all temporal indices downstairs. The extrinsic curvature of constant-$t$ hyper-surfaces, for our metric, is given by $K_{\mu\nu} = \Gamma^{0}_{\mu\nu}$ with non-zero components $K_{ij} = \Gamma_{ij}^{0} = \dot{g}_{ij} /2$. Throughout we will actually work with the perturbed extrinsic curvature defined by $\delta K_{\mu\nu} = K_{\mu\nu} - H h_{\mu\nu}$, where $h_{\mu\nu} = g_{\mu\nu}+\delta^0_\mu\delta^0_\nu$. \\

\noindent Now it is well-known that the full 20 components of the Riemann tensor are not independent of the extrinsic curvature. Indeed, only the three-dimensional part of the full four-dimensional object is independent and we denote this object by $\tilde{R}_{ijkl}$\footnote{Throughout we use a tilde to represent three-dimensional objects.} (see e.g. \cite{Gourgoulhon:2007ue}). A further simplification comes from the fact that the Weyl tensor is identically zero in three dimensions, so the three-dimensional Riemann tensor is completely determined by the three-dimensional Ricci tensor $\tilde{R}_{ij}$. This is most easily seen by counting the number of degrees of freedom of the Weyl tensor. The Riemann tensor in $D$ dimensions has $\frac{1}{12}D^2(D^2-1)$ degrees of freedom, while its trace has $\frac{1}{2}D(D+1)$ degrees of freedom. The Weyl tensor is the traceless part of the Riemann tensor, so in three dimensions it has $6-6=0$. We can also use only the spatial components of $\delta K_{\mu\nu}$ without loss of generality. Our general action is therefore
\begin{eqnarray}
S = S_{0} +  \int d^4 x \sqrt{-g} \, F(\tilde{R}_{ij}, \delta K_{ij},\nabla_{0}, \nabla_{i}),
\end{eqnarray}
where we have separated out $S_{0}$ which includes the Einstein-Hilbert action plus the terms required to make the unperturbed metric a consistent solution. This part of the action describes the minimal set-up of slow-roll inflation with all other operators describing higher derivative corrections. Any operators not contained in $S_{0}$ start at quadratic order in $\gamma$ so do not affect the tadpole cancellation \cite{Cheung:2007st}: they capture all different theories of cosmological perturbations on the same FRW background. Indeed, $\delta K_{ij}$ is a perturbed object by construction and 
$\tilde{R}_{ij}$ vanishes on the background. 
We have \cite{Cheung:2007st}
\begin{align}
S_{0} = \frac{M_{\text{pl}}^2}{2} \int d^4 x \sqrt{-g} \,\big[R - 2 (\dot{H} + 3H^2) + 2\dot{H} g^{00}\big],
\end{align}
and we remind the reader that we are working in the limit where $H$ and $\dot{H}$ do not vary significantly in one Hubble time: all time dependence of the action is slow-roll suppressed. We note that the perturbed action $S-S_{0}$ is derivatively coupled: in the general EFToI the only terms without derivatives are polynomials in $g^{00}$ and for us these are all trivial. \\

\noindent There is one final simplification we can make before moving onto field redefinitions: since $\tilde{R}_{ij}$ and $\delta K_{ij}$ are three-dimensional objects we only need to use the three-dimensional covariant derivative $\tilde{\nabla}_{i}$. Indeed, out of the non-vanishing Christoffel symbols only the three-dimensional one $\tilde{\Gamma}^{i}_{jk}$ cannot be expressed in terms of the extrinsic curvature. We can also treat all temporal covariant derivatives as partial ones $\partial_{t}$ for the same reason. We therefore have
\begin{eqnarray}
S = S_{0} +  \int d^4 x \sqrt{-g} \, F(\tilde{R}_{ij}, \delta K_{ij},\partial_{t}, \tilde{\nabla}_{i}).
\end{eqnarray}
Of course in some cases it will be wise to use covariant derivatives and only make this final simplification when we come to expand the action. 

\subsection{Eliminating the three-dimensional Ricci tensor}
We will now show that to construct graviton bispectra it is sufficient to work with the restricted action that does not depend on $\tilde{R}_{ij}$. Both $\tilde{R}_{ij}$ and $\delta K_{ij}$ start at linear order in perturbations so the graviton action up to cubic order, that comes in addition to the Einstein-Hilbert part, comes from EFToI operators that are at most cubic in these building blocks. \\

\noindent First consider operators that are constructed out of three building blocks i.e. are of the schematic form: $\tilde{R}^3$, $\tilde{R}^2 \delta K$, $\tilde{R} \delta K^2$ and $\delta K^3$ where we have suppressed indices and derivatives. Above we have argued that we can always use partial time derivatives, while for spatial derivatives the difference between using a partial one and a covariant one is captured by $\tilde{\Gamma}^{i}_{jk}$ which starts at linear order in perturbations. It follows that the difference between using a partial spatial derivative and a three-dimensional covariant one is $\mathcal{O}(\gamma^2)$, which for three building block operators will only introduce differences at $\mathcal{O}(\gamma^4)$. For our interests we can therefore treat all derivatives as partial ones for three building block operators. Now such operators do not contribute to the quadratic action for the graviton so to compute the cubic wavefunction coefficient we only need to consider contact diagrams where all external lines are on-shell in which case any appearances of $\partial^2 \gamma_{ij}$ are degenerate with $\dot{\gamma}_{ij}$, and its time derivatives, by the graviton equation of motion:
\begin{align} \label{GravitonEOM}
\ddot{\gamma}_{ij} + 3H \dot{\gamma}_{ij} - a^{-2} \partial^2 \gamma_{ij} = 0.
\end{align}
As we will discuss below, the two-derivative quadratic action can be brought into the Einstein-Hilbert form without loss of generality \cite{Creminelli:2014wna}, with higher-derivative corrections that we treat perturbatively such that \eqref{GravitonEOM} is always the on-shell relation. Once we remove all copies of $\partial^2 \gamma_{ij}$, all remaining interactions can be derived from $\delta K_{ij}$ and its derivatives since at linear order we have $\delta K_{ij} \sim \dot{\gamma}_{ij}$. So for three building blocks operators we can safely ignore operators containing $\tilde{R}_{ij} \sim \partial^2 \gamma_{ij}$, at least when computing the on-shell cubic vertices. We can also make the redundancy of any operators containing $\tilde{R}_{ij}$ manifest with field redefinitions, as we will show below. We provide more information about the contact diagrams we will be computing in Section \ref{Generalities}, and provide the general on-shell action coming from three building block operators in Section \ref{GravitonInteractions}. \\

\noindent Two and single building block operators are slightly more complicated but nevertheless we can still reduce the action to one constructed from $\delta K_{ij}$ only using field redefinitions. Let's start by considering which operators can contribute to the action up to cubic order. Since the graviton is transverse all vector quantities start at quadratic order in perturbations: $\nabla_{i} \tilde{R}_{ij} \sim \mathcal{O}(\gamma^2)$ and $\nabla_{i} \delta K_{ij} \sim \mathcal{O}(\gamma^2)$. It follows that all spatial covariant derivatives must be contracted with other spatial covariant derivatives since any other scalar quantities will start at quartic order in perturbations. The action for two building block operators is therefore
\begin{align} \label{TwoBB}
S = S_{0} +  M_{\text{pl}}^2 \int d^4 x \sqrt{-g} \,\left( \delta K^{ij} \mathcal{O}_{(0)} \delta K_{ij} +  \tilde{R}^{ij} \mathcal{O}_{(1)} \delta K_{ij} + \tilde{R}^{ij} \mathcal{O}_{(2)} \tilde{R}_{ij} \right),
\end{align}  
where $\mathcal{O}_{(i)}$ are derivative operators constructed out of $\nabla_{0}$ and $\tilde{\nabla}^2 = \tilde{\nabla}^{i} \tilde{\nabla}_{i}$, and we have integrated by parts to move all derivatives onto a single building block. One might worry about boundary terms that could affect the wavefunction, but in Section \ref{Generalities} we will show that there are always enough derivatives for the boundary terms to vanish. For $\mathcal{O}_{(i)}$, $i$ counts its negative mass dimension which we can fix using $M_{\text{pl}}$:
\begin{align}
\mathcal{O}_{(i)} = \frac{1}{M_{\text{pl}}^i} \sum_{m,n} b_{m,n} \nabla_{0}^{m} \tilde{\nabla}^{2n},
\end{align}
where $b_{m,n}$ are constant couplings with mass dimension $-(m+2n)$ (recall that we are working with a dimensionless $\gamma_{ij}$). Note that we have not included any terms that depend on the trace of the extrinsic curvature since this object vanishes to all orders in $\gamma$\footnote{This can be most easily seen by writing $2 K = g^{ij}\dot{g}_{ij} = g^{-1} \dot{g}$ where $g$ is the determinant of the spatial metric $g_{ij}$. Since the graviton is traceless, it drops out of $g$ and therefore drops out of the trace of the extrinsic curvature.}, and we have not included any terms that depend on the three-dimensional Ricci scalar $\tilde{R}$ since a linear term can be eliminated by Gauss-Codazzi relations \cite{Creminelli:2014wna}\footnote{Eliminating a linear term in $\tilde{R}$ is more involved if we allow for time-dependent couplings but it can still be done \cite{Creminelli:2014wna,Bordin:2017hal}.}, while non-linear terms start at $\mathcal{O}(\gamma^4)$ since $\tilde{R} \sim \mathcal{O}(\gamma^2)$. \\

\noindent Now it was shown in \cite{Creminelli:2014wna} that we can set the coefficient of $\delta K^{ij} \delta K_{ij}$ to any value we like such that that the quadratic action takes the canonical form. This is achieved thanks to conformal and disformal transformations of the metric. We can also go further by considering additional field redefinitions of the spatial components of the metric of the form
\begin{align} 
\delta g^{ij} =\mathcal{O}_{(2)} \tilde{R}^{ij} + \mathcal{O}_{(1)} \delta K^{ij}. \label{FieldRedefinitions2}
\end{align}
Note that different copies of $\mathcal{O}_{i}$ with the same $i$ are not necessarily the same derivative operators. The linear variation of the minimal action can be written in terms of the EFToI operators $\tilde{R}_{ij}$ and $\delta K_{ij}$. We find
\begin{align} \label{MinimalVariation}
\delta S_{0} = \frac{M_{\text{pl}}^2}{2} \int d^4 x \sqrt{-g}\,\big[\tilde{G}_{ij} + \nabla_{0} \delta K_{ij} + 3 H \delta K_{ij}\big] \delta g^{ij},
\end{align} 
where in computing the variation of the Einstein-Hilbert action we have dropped contributions that depend on $g^{ij}\delta K_{ij}$ since as we mentioned above this object vanishes to all orders in $\gamma$, and we have dropped two building block terms since under our chosen field redefinition these will only correct three building block operators which we have already considered. The three-dimensional Einstein tensor is given by $\tilde{G}_{ij} = \tilde{R}_{ij} - \frac{1}{2}g_{ij} \tilde{R}$. One might wonder why we have only transformed $\delta S_{0}$: this is simply because we are treating all two building block operators as small corrections to the Einstein-Hilbert action. We have done this such that we can use the usual on-shell condition \eqref{GravitonEOM}. This means that performing the field redefinition on two building block operators can only generate even further suppressed operators. We will discuss perturbation theory in more detail in Section \ref{Generalities}. \\

\noindent Now from these expressions it is clear that, up to three building block operators which we have already considered, we can eliminate all $\tilde{R}^{ij} \mathcal{O}_{(1)} \delta K_{ij}$ and $\tilde{R}^{ij} \mathcal{O}_{(2)} \tilde{R}_{ij} $ operators from \eqref{TwoBB}. Indeed we can first use the product $\tilde{G}_{ij} \mathcal{O}_{(2)} \tilde{R}^{ij}$ in \eqref{MinimalVariation} to eliminate $\tilde{R}^{ij} \mathcal{O}_{(2)} \tilde{R}_{ij}$ operators while renormalising $\tilde{R}^{ij} \mathcal{O}_{(1)} \delta K_{ij}$ and introducing non-linear terms in $\tilde{R}$ which don't contribute to the action up to cubic order. We can then use the product $\tilde{G}_{ij} \mathcal{O}_{(1)} \delta K^{ij}$ to eliminate all $\tilde{R}^{ij} \mathcal{O}_{(1)} \delta K_{ij}$ operators while renormalising $\delta K^{ij} \mathcal{O}_{(0)} \delta K_{ij}$ and introducing other operators that don't contribute to the cubic action. We cannot also eliminate the $\delta K^{ij} \mathcal{O}_{(0)} \delta K_{ij}$ operators since there is not enough freedom in \eqref{FieldRedefinitions2}. Note that in terms of $\gamma_{ij}$, these field redefinitions shift $\gamma_{ij}$ by terms with derivatives and then scale invariance ensures that the redefinitions vanish at late-times. So wavefunction coefficients and cosmological correlators evaluated at the end of inflation are not affected by these field redefinitions. \\

\noindent Now it is clear how we can generalise this discussion to manifestly remove all copies of $\tilde{R}_{ij}$ from three building block operators. We simply need to use field redefinitions of the schematic form $\delta g = \tilde{R}^2 + \tilde{R} \delta K + \delta K \delta K$ where we have suppressed indices and derivatives. Acting on the Einstein-Hilbert action with these field redefinitions allows us to reduce three building block operators to ones cubic in $\delta K_{ij}$. It follows that the most general action that can contribute to graviton bispectra at tree-level is 
\begin{align} \label{AfterRedefinitions}
S = S_{0} +  M_{\text{pl}}^2 \int d^4 x \sqrt{-g}\,\big[ \delta K^{ij} \mathcal{O}_{(0)} \delta K_{ij} + \mathcal{O}(\delta K ^3)\big].
\end{align} 
For reasons that will become clear a bit later on we keep the coefficient of $\delta K^{ij} \delta K_{ij}$ non-zero for now. \\

\noindent We have therefore shown in this section that to compute the bispectra for gravitons, in addition to the minimal action, we only need to consider EFToI actions constructed out of the extrinsic curvature. Time derivatives can always be taken to be partial ones and spatial derivatives only need to be covariant for two building block operators and they only need to be contracted amongst themselves.


\section{A general action to cubic order} \label{GravitonInteractions}
We can now compute the general action up to cubic order in $\gamma_{ij}$. The Einstein-Hilbert part of the action yields
\begin{align}
\label{GR_action}
S_{\gamma, \text{GR}} = \frac{M_{\text{pl}}^2}{8} \int dt d^{3}x a^3(t) [\dot{\gamma}_{ij} \dot{\gamma}_{ij} - a^{-2} \partial_{k} \gamma_{ij} \partial_{k} \gamma_{ij} + a^{-2}(2 \gamma_{ik}\gamma_{jl} - \gamma_{ij}\gamma_{kl}) \partial_{k}\partial_{l} \gamma_{ij}] + \mathcal{O}(\gamma^4),
\end{align}
and we remind the reader that for $\gamma_{ij}$ operators, spatial indices are raised and lowered with $\delta_{ij}$ so we will not be so careful about the placement of indices. For the corrections to the minimal action we consider the two and three building block operators separately.

\subsection{Two building block operators}

\noindent Let's start with operators quadratic in $\delta K_{ij}$ which take the form
\begin{align} \nn
S &=  M_{\text{pl}}^2 \int d^4 x \sqrt{-g} \,\delta K^{ij} \mathcal{O}_{(0)} \delta K_{ij}  \\
 &= M_{\text{pl}}^2 \int d^4 x \sqrt{-g} \,\sum_{m,n} g_{n,m} \nabla_{0}^{2n} \delta K^{ij}  \tilde{\nabla}^{2m} \delta K_{ij},
 \label{eq:ActionSec3}
\end{align}
where we have taken the number of time derivatives to be even such that this operator is not a total derivative, and we integrated by parts to separate the time and space derivatives. One might worry about the ordering of the covariant derivatives, but we can guarantee this ordering up to curvature corrections which we have already dealt with above. The couplings $g_{m,n}$ are in general dimensionful. Again, in Section \ref{Generalities} we will show that any boundary terms that might arise when we integrate by parts will not contribute to the late-time wavefunction. As we explained above we can also safely take time derivatives to be partial ones - the difference can be accounted for by adding appropriate three building block operators - and we will denote the $n^{\text{th}}$ partial time derivative by $(\delta K_{ij})^{n}$. Note that this procedure does not affect the couplings $g_{n,m}$ and only changes the couplings of three building block operators. We will construct the most general set of bispectra so we can do this without loss of generality. We would now like to expand the redefined action
\be
S = M_{\text{pl}}^2 \int d^4 x \sqrt{-g} \,\sum_{m,n} g_{n,m} \partial_{0}^{2n} \delta K^{ij}  \tilde{\nabla}^{2m} \delta K_{ij},
 \label{eq:ActionSec3B}
\ee
to find the graviton action up to cubic order which for these two building block operators we denote by $S_{\gamma, \text{2BB}}$. Recall that the metric determinant is independent of $\gamma$, while $\delta K$ vanishes on the background. So to find the action up to cubic order we always need to expand one of the $\delta K$ to linear order and the other to quadratic order. \\

\noindent First consider $m=0$ where all derivatives are temporal ones. In this case we see that by expanding the action there are no corrections at $\mathcal{O}(\gamma^3)$. For $n=0$ this was noticed in \cite{Maldacena:2002vr} and it holds true for all $n$. This can be most easily seen by writing the operators as $(\delta K^{i}{}_{j})^{2n}  \delta K^{j}{}_{i}$ and by noticing that, for all $n$, $(\delta K^{i}{}_{j})^{2n}$ is \textit{symmetric} at linear order while \textit{anti-symmetric} at quadratic order\footnote{Of course we should raise and lower these indices before we replace covariant derivatives with partial ones. }. We have
\begin{align} \label{eq:EqForKij}
\delta K^{i}{}_{j} = \frac{1}{2} a^{2}(t) g^{il} \partial_{t} (e^{\gamma})_{lj} =  \frac{1}{2} \dot{\gamma}^{i}{}_{j} + \frac{1}{4}(\dot{\gamma}^{i}{}_{l} \gamma_{lj} - \gamma^{i}{}_{l} \dot{\gamma}_{lj}) + \mathcal{O}(\gamma^3),
\end{align}
and the symmetry properties of this object are not altered when we take the partial derivative. The action for the graviton perturbation, for $m=0$, is then simply the quadratic one given by
\begin{align}
S_{\gamma, \text{2BB}} \supset \frac{M_{\text{pl}}^2}{4} \int dt d^3 x a^3(t) \sum_{n} g_{n,0} (\gamma_{ij})^{2n+1} \dot{\gamma}_{ij}.
\end{align}
As we mentioned above, for $n=0$ it looks like the two-derivative action is altered by this operator but we can always set $g_{0,0} = 0$ using field redefinitions \cite{Creminelli:2014wna} (we will comment on this further below). \\

\noindent Now for $m \neq 0$ we need to compute the spatial covariant derivatives of the extrinsic curvature, and in this case we cannot simply treat them as partial derivatives since $\tilde{\Gamma}^{i}_{jk}$ cannot be written in terms of the extrinsic curvature. As we showed above, $(\delta K^{i}{}_{j})^{2n}$ is anti-symmetric at quadratic order, while at linear order we can treat covariant derivatives as partial ones and we have
\begin{align}
\tilde{\nabla}^{2m} \delta K^{i}{}_{j} = \frac{1}{2}a^{-2m} \partial^{2m} \dot{\gamma}^{i}{}_{j} + \mathcal{O}(\gamma^2),
\end{align}
which is symmetric. It follows that the only contribution to the cubic action comes from taking $(\delta K^{i}{}_{j})^{2n}$ at linear order and $\tilde{\nabla}^{2m} \delta K^{i}{}_{j}$ at quadratic order, and only keeping the symmetric contributions. Let's now focus on computing $\tilde{\nabla}^{2m} \delta K^{i}{}_{j}$. For $m=1$ we find, by direct computation, that up to quadratic order in $\gamma_{ij}$ we have
\begin{align} \label{SpatialCovariant}
\tilde{\nabla}^2 \delta K^{i}{}_{j} = \frac{1}{2}a^{-2}\left( \partial^2 \dot{\gamma}^{i}{}_{j} - \gamma_{lm} \partial_{l}\partial_{m} \dot{\gamma}^{i}{}_{j} + Q_{s}[\gamma, \dot{\gamma}]^{i}{}_{j}\right) + \text{anti-symmetric},
\end{align}
where the symmetric function $Q_{s}$ is given by 
\begin{align}
Q_{s}[\gamma, \dot{\gamma}]_{ij}  = - \partial_{i} \gamma_{kl} \partial_{k} \dot{\gamma}_{lj} + \partial_{l} \gamma_{jk} \partial_{k} \dot{\gamma}_{il} + (i \leftrightarrow j),
\end{align}
and we leave implicit the $\mathcal{O}(\gamma^2)$ anti-symmetric contributions to $\tilde{\nabla}^2 \delta K^{i}{}_{j}$ since when contracted with the linear expansion of $(\delta K^{i}{}_{j})^{2n}$ they will sum to zero. The first term in \eqref{SpatialCovariant} comes from treating both covariant derivatives as partial ones, the second comes from expanding the metric that contracts the two spatial derivatives, while the final structure comes from the Christoffel symbols.  With this expression at hand we can now recursively generalise to any $m \neq 0$. Again working up to quadratic order, we find
  \begin{align}
\tilde{\nabla}^{2m} \delta K^{i}{}_{j} = \frac{1}{2}a^{-2m} \left(   \partial^{2m} \dot{\gamma}^{i}{}_{j}  - m\gamma_{lk} \partial^{2m-2} \partial_{l}\partial_{k}  \dot{\gamma}^{i}{}_{j} + \sum_{p}^{m-1} \partial^{2p} Q_{s}[\gamma, \partial^{2m-2 - 2p} \dot{\gamma}]^{i}{}_{j} \right), 
 \end{align}
 where again we have dropped anti-symmetric terms. The action up to cubic order, now with $m=0$ included, is therefore
 \begin{align} \label{2BB}
&S_{\gamma, \text{2BB}} = \frac{M_{\text{pl}}^2}{4} \int dt d^3 x a^3(t) \sum_{n} g_{n,0} (\gamma_{ij})^{2n+1} \dot{\gamma}_{ij} ~  + \nonumber \\ \frac{M_{\text{pl}}^2}{4} \int dt d^3 x a^{3-2m}(t) & \sum_{n,m=1} g_{n,m} (\gamma_{ij})^{2n+1}\left(   \partial^{2m} \dot{\gamma}_{ij}  - m\gamma^{lk} \partial^{2m-2} \partial_{l}\partial_{k}  \dot{\gamma}_{ij} + \sum_{p}^{m-1} \partial^{2p} Q_{s}[\gamma, \partial^{2m-2 - 2p} \dot{\gamma}]_{ij} \right).
 \end{align}
This is the most general action, from two building block operators, that can contribute to the graviton bispectra in the EFToI. It contains a sum of quadratic and cubic terms with their relative coefficients tied together by the linear realization of spatial diffeomorphisms and nonlinear realization of time diffeomorphisms. Note that in this section we have freely added and subtracted three building block operators. We were allowed to do this, because we are going to consider these operators in their full generality in the next section.
\subsection{Three building block operators}
Now consider three building block operators which all start at $\mathcal{O}(\gamma^3)$. As we discussed above we can treat all covariant derivatives as partial ones and since $\delta K_{ij} \sim \dot{\gamma}_{ij}$ at linear order, we simply need to write down all independent contractions of three copies of $\dot{\gamma}_{ij}$, and spatial derivatives. This problem was tackled in \cite{Cabass:2021fnw} at the level of polarisation tensors and spatial momenta but it is simple to convert it into a Lagrangian statement. There are five independent tensor structures once we use the fact that the graviton is transverse and traceless, and integrate by parts (which is equivalent to momentum conservation). These structures are organised by the number of spatial derivatives that are contracted with a graviton and we denote this number by $\alpha$. For parity-even interactions this is an even number and for $\alpha=0,4,6$ there is a single structure, while for $\alpha=2$ there are two. We cannot have $\alpha \geq 8$ since there are only $6$ free indices across the three gravitons.  For each $\gamma_{ij}$ we can add additional time derivatives, while additional spatial derivatives can be restricted to two fields only by integration by parts, and two derivatives that are contracted with each other should act on different fields. Any other contractions can be removed by the graviton's equation of motion in favour of time derivatives which we are already adding. We have
\begin{align}
S_{\gamma, \text{3BB}} = \sum_{\alpha} \int dt d^3 x  \mathcal{L}_{\gamma,\alpha},
\end{align}
where 
\begin{align}
\mathcal{L}_{\gamma,\alpha=0} &=  \sum_{n_{1},n_{2},n_{3},p} a^{q}(t)  h^{0}_{n_{1},n_{2},n_{3},p} (\gamma_{ij})^{1+n_{1}} \partial_{i_{1} \ldots i_{p}} (\gamma_{jk})^{1+n_{2}} \partial_{i_{1} \ldots i_{p}} (\gamma_{ki})^{1+n_{3}} , \\
\mathcal{L}_{\gamma,\alpha=2} &=  \sum_{n_{1},n_{2},n_{3},p} a^{q}(t)  h^{2}_{n_{1},n_{2},n_{3},p} (\gamma_{ij})^{1+n_{1}} \partial_{i_{1} \ldots i_{p}} \partial_{i} (\gamma_{lm})^{1+n_{2}} \partial_{i_{1} \ldots i_{p}} \partial_{j} (\gamma_{lm})^{1+n_{3}}  \\
 &  +  \sum_{n_{1},n_{2},n_{3},p}  a^{q}(t) \hat{h}^{2}_{n_{1},n_{2},n_{3},p}  (\gamma_{ij})^{1+n_{1}} \partial_{i_{1} \ldots i_{p}} (\gamma_{lm})^{1+n_{2}} \partial_{i_{1} \ldots i_{p}} \partial_{i} \partial_{l} (\gamma_{jm})^{1+n_{3}} , \\
\mathcal{L}_{\gamma,\alpha=4} &=  \sum_{n_{1},n_{2},n_{3},p} a^{q}(t)  h^{4}_{n_{1},n_{2},n_{3},p} \partial_{i} (\gamma_{lk})^{1+n_{1}} \partial_{i_{1} \ldots i_{p}} \partial_{j} (\gamma_{mk})^{1+n_{2}} \partial_{i_{1} \ldots i_{p}} \partial_{l}\partial_{m} (\gamma_{ij})^{1+n_{3}} , \\
\mathcal{L}_{\gamma,\alpha=6} &=   \sum_{n_{1},n_{2},n_{3},p} a^{q}(t)  h^{6}_{n_{1},n_{2},n_{3},p} \partial_{m}\partial_{k} (\gamma_{il})^{1+n_{1}} \partial_{i_{1} \ldots i_{p}} \partial_{i}\partial_{n} (\gamma_{jm})^{1+n_{2}} \partial_{i_{1} \ldots i_{p}} \partial_{j}\partial_{l} (\gamma_{kn})^{1+n_{3}},
\end{align}
with constant couplings $h^{\alpha}_{n_{1},n_{2},n_{3},p}$ and $q = 3 -2p -\alpha$ which is fixed by scale invariance. Despite the complicated looking nature of these interactions, in Section \ref{TypeII} we will show that the resulting bispectra take a very compact form\footnote{When computing the bispectra we will not actually use these explicit Lagrangian expressions, instead for three-building block operators we will use the MLT \cite{MLT} to efficiently write down all possibilities.}. \\

\noindent We have therefore written down the most general actions that can contribute to graviton bispectra, $S_{\gamma, \text{2BB}}$ and $S_{\gamma, \text{3BB}}$. In the following sections we discuss the diagrams that contribute to the cubic wavefunction and present compact expressions for the bispectra, but first we close this section by converting our actions to conformal time and addressing the fact that these general actions can contain higher-order time derivatives. 

\subsection{Converting to conformal time}

Throughout we have been working in cosmological time but when we come to compute the late-time wavefunction we would like to do so in conformal time $\eta$ where we evolve perturbations from the far past at $\eta = - \infty$ to the boundary of quasi-de Sitter space, or the end of inflation, at $\eta=0$. We therefore need to change coordinates in the actions we have just derived so that the background metric is 
\begin{align}
ds^2 = a^2 (\eta)[- d \eta^2 + d x^2],
\end{align}
and we will approximate the scale factor as $a(\eta) = -1 / (H \eta)$ which is the de Sitter one in Poincar\'{e} or flat-slicing coordinates. \\

\noindent First consider two building block operators and the action \eqref{2BB}. When we convert an object of the form $(\gamma_{ij})^{n}$ to conformal time we will generate a sum of terms with all derivatives from $1$ to $n$. For $m=0$, we can use the fact that the couplings $g_{n,0}$ are arbitrary, to simply replace all cosmological time derivatives with conformal ones without loss of generality, while introducing the required scale factors. We can also now make use our freedom to fix the coefficient of $\delta K^{ij} \delta K_{ij}$ to anything we like to eliminate any quadratic terms with two derivatives. This is the statement that the graviton speed of sound can always be fixed to unity \cite{Creminelli:2014wna}, and as expected this statement is true in both cosmological and conformal time\footnote{In fact, we can also bring the two derivative action to its canonical GR form by redefining time, then redefining $\gamma_{ij}$, then finally redefining all couplings. We can always do this since we only have one field in the theory.}. This procedure will generate terms with an overall odd number of time derivatives but these can always be integrated by parts in favour of ones with an even number of derivatives, and as always we can safely drop any boundary terms that may arise (see Section \ref{Generalities}). We can therefore write the first line of \eqref{2BB} as
\begin{align} \label{2BBCTZeroM}
S_{\gamma, \text{2BB}} \supset \frac{M_{\text{pl}}^2}{4} \int d \eta d^3 x  \sum_{n=1} g_{n,0} a^{2-2n}(\eta) (\gamma_{ij})^{2n+1} \gamma'_{ij},
\end{align}
where we keep the same labels for the arbitrary couplings and still use brackets to denote higher-order time derivatives, with the integration variables informing the reader which coordinate it refers to. We note that the couplings are now not exactly as they were defined in the covariant action, instead these new couplings are linear sums of the old ones. We now start the sum from $n=1$ since we have at least four time derivatives, without loss of generality, and importantly adding $\delta K^{ij} \delta K_{ij}$ does not introduce any additional terms since it does not contribute to the cubic action for $\gamma_{ij}$. The scale symmetry of the action is now crystal clear in these coordinates and offers a good consistency check that we have the correct number of scale factors: we need four scale factors to cancel the scaling of $d \eta d^3 x$ under a scale transformation, then we need to remove a scale factor for each derivative. For an action like the one above with $2n+2$ derivatives we therefore need $4-(2n+2) = 2-2n$ scale factors. \\

\noindent We can deal with the second line of \eqref{2BB} in a similar way. The structure of the time derivatives is the same for each term so we can simply replace all cosmological time derivatives with conformal ones, while adding the appropriate number of scale factors, and redefining the couplings $g_{m,n}$. Here there is no need to add operators to remove certain terms so the sums still run from $n=0$ and $m=1$. However, now that we have cubic interactions we cannot simply drop all terms with an overall odd number of time derivatives, or equivalently those with an even number that come from transforming $(\gamma_{ij})^{2n+1}$ to conformal time. So for $m \neq 0$ we must replace $2n$ with $n$. In conformal time the full \eqref{2BB} then becomes
\begin{align} \label{2BBCT}
&S_{\gamma, \text{2BB}} = \frac{M_{\text{pl}}^2}{4} \int d \eta d^3 x  \sum_{n=1} g_{n,0} a^{2-2n}(\eta) (\gamma_{ij})^{2n+1} \gamma'_{ij} ~  + \nonumber \\ \frac{M_{\text{pl}}^2}{4} \int d \eta d^3 x  & \sum_{n=0,m=1} g_{n,m} a^{q}(\eta) (\gamma_{ij})^{n+1}\left(   \partial^{2m} \gamma'_{ij}  - m\gamma^{lk} \partial^{2m-2} \partial_{l}\partial_{k} \gamma'_{ij} + \sum_{p}^{m-1} \partial^{2p} Q_{s}[\gamma, \partial^{2m-2 - 2p} \gamma']_{ij} \right),
\end{align}
where $q = 2 -2m -n$. Now each operator has at least four derivatives meaning that it is impossible to integrate by parts to generate operators with fewer derivatives. This can only happen when a time derivative acts on a scale factor, but when there are exactly four derivatives there are no scale factors. This will have important consequences for the structure of the corresponding bispectra. \\

\noindent For the three building block operators we use the same logic such that the vertices take exactly the same structure as before, but now with a different number of scale factors. We now have
\begin{align}
S_{\gamma, \text{3BB}} = \sum_{\alpha} \int d \eta d^3 x  \mathcal{L}_{\gamma,\alpha},
\end{align}
where
\begin{align} \label{3BBCT}
\mathcal{L}_{\gamma,\alpha=0} &=  \sum_{n_{1},n_{2},n_{3},p} a^{r}(\eta) h^{0}_{n_{1},n_{2},n_{3},p} (\gamma_{ij})^{1+n_{1}} \partial_{i_{1} \ldots i_{p}} (\gamma_{jk})^{1+n_{2}} \partial_{i_{1} \ldots i_{p}} (\gamma_{ki})^{1+n_{3}} , \\
\mathcal{L}_{\gamma,\alpha=2} &=   \sum_{n_{1},n_{2},n_{3},p} a^{r}(\eta) h^{2}_{n_{1},n_{2},n_{3},p} (\gamma_{ij})^{1+n_{1}} \partial_{i_{1} \ldots i_{p}} \partial_{i} (\gamma_{lm})^{1+n_{2}} \partial_{i_{1} \ldots i_{p}} \partial_{j} (\gamma_{lm})^{1+n_{3}}  \\
&  +  \sum_{n_{1},n_{2},n_{3},p} a^{r}(\eta) \hat{h}^{2}_{n_{1},n_{2},n_{3},p} (\gamma_{ij})^{1+n_{1}} \partial_{i_{1} \ldots i_{p}} (\gamma_{lm})^{1+n_{2}} \partial_{i_{1} \ldots i_{p}} \partial_{i} \partial_{l} (\gamma_{jm})^{1+n_{3}} , \\
\mathcal{L}_{\gamma,\alpha=4} &= \sum_{n_{1},n_{2},n_{3},p} a^{r}(\eta) h^{4}_{n_{1},n_{2},n_{3},p} \partial_{i} (\gamma_{lk})^{1+n_{1}} \partial_{i_{1} \ldots i_{p}} \partial_{j} (\gamma_{mk})^{1+n_{2}} \partial_{i_{1} \ldots i_{p}} \partial_{l}\partial_{m} (\gamma_{ij})^{1+n_{3}} , \\
\mathcal{L}_{\gamma,\alpha=6} &=  \sum_{n_{1},n_{2},n_{3},p} a^{r}(\eta) h^{6}_{n_{1},n_{2},n_{3},p} \partial_{m}\partial_{k} (\gamma_{il})^{1+n_{1}} \partial_{i_{1} \ldots i_{p}} \partial_{i}\partial_{n} (\gamma_{jm})^{1+n_{2}} \partial_{i_{1} \ldots i_{p}} \partial_{j}\partial_{l} (\gamma_{kn})^{1+n_{3}},
\end{align}
and we have defined $r = 1-2p-n_{1}-n_{2}-n_{3}-\alpha$, and have redefined the couplings. These are the actions we will use to compute the graviton bispectra.

\subsection{A comment on higher-order time derivatives}
Our actions for both two and three building block operators can contain higher-order time derivatives which may concern the reader. When treated perturbatively such interactions should be harmless, however there are other actions that give exactly the same bispectra as the ones we are going to compute which have at most one time derivative per field. This should reassure a concerned reader. We use the actions above since they have a very simple covariant form, and the properties of $\delta K^{i}{}_{j}$ made expanding the action a relatively simple task. \\

\noindent To see how we could remove higher-order time derivatives lets go back to the field redefinitions we used to eliminate the $\tilde{R}^2$ and $\tilde{R} \delta K$ operators (again we are dropping indices and derivatives). We have three different structures in the action and two different structures in the field redefinition which is schematically $\delta g = \tilde{R} + \delta  K$. So instead of removing the $\tilde{R}^2$ and $\tilde{R} \delta K$ operators leaving us with only $\delta K^2$ ones, we could remove the $\delta K^2$ ones and use the remaining freedom to fix the tuning between $\tilde{R}^2$ and $\tilde{R} \delta K$. Such tunings can be fixed to remove all higher-order time derivatives in the quadratic corrections to the graviton action. For example, consider the following operator:
\begin{align}
\tilde{R}^{ij} \tilde{\nabla}^{2m} \nabla_{0}^{2n} \tilde{R}_{ij}  \supset \partial^2 \gamma_{ij} \partial^{2m+2} (\gamma_{ij})^{2n},
\end{align}
which contains higher-order time derivatives that cannot be integrated by parts for $n \geq 2$. We can cancel these using the operators: 
\begin{align}
\tilde{R}^{ij} \tilde{\nabla}^{2m+2} \nabla_{0}^{2n-1} \delta K_{ij}  \supset \partial^2 \gamma_{ij} \partial^{2m+2} (\gamma_{ij})^{2n},
\end{align}
since their coefficients can be chosen freely thanks to the field redefinitions at our disposal. The resulting action would still contain quadratic corrections but these would only come from $\tilde{R}^{ij} \tilde{\nabla}^{2m} \tilde{R}_{ij}$ and so would only contain spatial derivatives. The cubic action is likely to take a more complicated form than the one we have above, however. Let us emphasise again that field redefinitions of this form do not change the wavefunction coefficients, and therefore do not change correlators. For these two building block operators it could still be the case that the resulting cubic vertices have higher-order time derivatives even though the quadratic ones do not. This is not an issue for the on-shell action as we can again use the equation of motion to eliminate them, but off-shell we would not be able to do this. At this stage we have exhausted all possible field redefinitions at the level of $g_{ij}$, which is the most sensible way to formulate these field redefinitions. So if the off-shell action was to also not have higher-order time derivatives, the tuning between the $\tilde{R}^2$ and $\tilde{R} \delta K$ operators that removes higher-order time derivatives at quadratic order must also be enough to remove them at cubic order.  \\

\noindent We can play the same game for three building block operators. At the level of the expanded action it is clear that we can use the equation of motion to remove all higher-order time derivatives but it can also be made manifest using covariant field redefinitions. For three building block operators we have $\tilde{R}^3$, $\tilde{R}^2 \delta K$, $\tilde{R} \delta K ^2$ and $\delta K^3$. We also have three different field redefinitions due to the three different bi-products of $\tilde{R}$ and $\delta K$. We can use this freedom to eliminate the $\tilde{R} \delta K ^2$ and $\delta K^3$ operators while tuning the coefficients between the $\tilde{R}^3$ and $\tilde{R}^2 \delta K$ ones such that there are no higher-order time derivatives. We choose to not eliminate these higher-order time derivatives since, as we will discuss in Section \ref{TypeII}, we can very easily extract the bispectra from \eqref{3BBCT} using the MLT. The bispectra are invariant under such field redefinitions so it makes sense to work with the action that lends itself to extracting correlators in the most efficient way.


\section{Generalities of contact and single exchange diagrams} \label{Generalities}

Before constructing EFToI bispectra, in this section we will provide more details on the diagrams that we need to compute. We will work within the WFU approach then extract expectation values using the usual formula. The WFU has been reviewed in many places in the recent literature e.g \cite{MLT,COT}, so here we skip such a review and concentrate on the diagrams of interest which are cubic contact diagrams and single exchange diagrams. The Feynman rules for computing these diagrams is reviewed in \cite{MLT,COT}, but we will briefly remind the reader of them in this section too. 

\subsection{Contact diagrams}
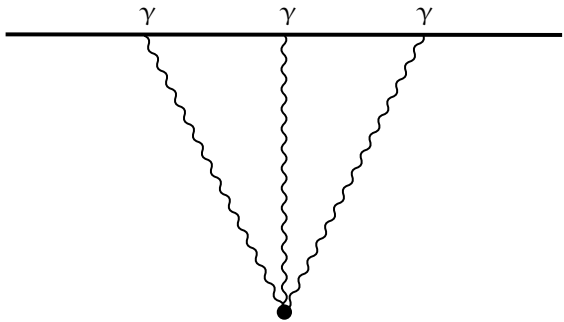
\begin{figure}
    \tikzset{every picture/.style={line width=0.75pt}} 
    
    \begin{tikzpicture}[x=0.75pt,y=0.75pt,yscale=-1,xscale=1]
    uncomment if require: \path (50,260); 
    
    \draw [line width=1.5]    (209.93,90.11) -- (490.6,90.4) ;
    \draw    (279.8,90.8) .. controls (282.04,91.53) and (282.79,93.02) .. (282.06,95.26) .. controls (281.33,97.5) and (282.08,98.99) .. (284.32,99.72) .. controls (286.56,100.45) and (287.31,101.94) .. (286.58,104.18) .. controls (285.85,106.42) and (286.61,107.91) .. (288.85,108.64) .. controls (291.09,109.37) and (291.84,110.86) .. (291.11,113.1) .. controls (290.38,115.34) and (291.13,116.83) .. (293.37,117.56) .. controls (295.61,118.29) and (296.36,119.78) .. (295.63,122.02) .. controls (294.9,124.26) and (295.65,125.74) .. (297.89,126.47) .. controls (300.13,127.2) and (300.88,128.69) .. (300.15,130.93) .. controls (299.42,133.17) and (300.17,134.66) .. (302.41,135.39) .. controls (304.65,136.12) and (305.41,137.61) .. (304.68,139.85) .. controls (303.95,142.09) and (304.7,143.58) .. (306.94,144.31) .. controls (309.18,145.04) and (309.93,146.53) .. (309.2,148.77) .. controls (308.47,151.01) and (309.22,152.5) .. (311.46,153.23) .. controls (313.7,153.96) and (314.45,155.45) .. (313.72,157.69) .. controls (312.99,159.93) and (313.74,161.42) .. (315.98,162.15) .. controls (318.22,162.88) and (318.98,164.37) .. (318.25,166.61) .. controls (317.52,168.85) and (318.27,170.34) .. (320.51,171.07) .. controls (322.75,171.8) and (323.5,173.29) .. (322.77,175.53) .. controls (322.04,177.77) and (322.79,179.26) .. (325.03,179.99) .. controls (327.27,180.72) and (328.02,182.21) .. (327.29,184.45) .. controls (326.56,186.69) and (327.31,188.18) .. (329.55,188.91) .. controls (331.79,189.64) and (332.54,191.12) .. (331.81,193.36) .. controls (331.08,195.6) and (331.84,197.09) .. (334.08,197.82) .. controls (336.32,198.55) and (337.07,200.04) .. (336.34,202.28) .. controls (335.61,204.52) and (336.36,206.01) .. (338.6,206.74) .. controls (340.84,207.47) and (341.59,208.96) .. (340.86,211.2) .. controls (340.13,213.44) and (340.88,214.93) .. (343.12,215.66) .. controls (345.36,216.39) and (346.11,217.88) .. (345.38,220.12) .. controls (344.65,222.36) and (345.4,223.85) .. (347.64,224.58) .. controls (349.88,225.31) and (350.64,226.8) .. (349.91,229.04) -- (350.53,230.28) -- (350.53,230.28) ;
    \draw    (420.6,90.4) .. controls (421.35,92.64) and (420.6,94.13) .. (418.36,94.87) .. controls (416.12,95.61) and (415.37,97.1) .. (416.12,99.34) .. controls (416.87,101.58) and (416.12,103.07) .. (413.88,103.81) .. controls (411.64,104.55) and (410.89,106.04) .. (411.64,108.28) .. controls (412.39,110.52) and (411.64,112.01) .. (409.4,112.75) .. controls (407.16,113.49) and (406.41,114.98) .. (407.16,117.22) .. controls (407.91,119.46) and (407.16,120.95) .. (404.92,121.69) .. controls (402.69,122.44) and (401.94,123.93) .. (402.69,126.16) .. controls (403.44,128.4) and (402.69,129.89) .. (400.45,130.63) .. controls (398.21,131.37) and (397.46,132.86) .. (398.21,135.1) .. controls (398.96,137.34) and (398.21,138.83) .. (395.97,139.58) .. controls (393.73,140.32) and (392.98,141.81) .. (393.73,144.05) .. controls (394.48,146.29) and (393.73,147.78) .. (391.49,148.52) .. controls (389.25,149.26) and (388.5,150.75) .. (389.25,152.99) .. controls (390,155.23) and (389.25,156.72) .. (387.01,157.46) .. controls (384.77,158.2) and (384.02,159.69) .. (384.77,161.93) .. controls (385.52,164.17) and (384.77,165.66) .. (382.53,166.4) .. controls (380.29,167.14) and (379.54,168.63) .. (380.29,170.87) .. controls (381.04,173.11) and (380.29,174.6) .. (378.05,175.34) .. controls (375.81,176.08) and (375.06,177.57) .. (375.81,179.81) .. controls (376.56,182.05) and (375.81,183.54) .. (373.57,184.28) .. controls (371.33,185.02) and (370.58,186.51) .. (371.33,188.75) .. controls (372.08,190.98) and (371.33,192.47) .. (369.1,193.22) .. controls (366.86,193.96) and (366.11,195.45) .. (366.86,197.69) .. controls (367.61,199.93) and (366.86,201.42) .. (364.62,202.16) .. controls (362.38,202.9) and (361.63,204.39) .. (362.38,206.63) .. controls (363.13,208.87) and (362.38,210.36) .. (360.14,211.1) .. controls (357.9,211.84) and (357.15,213.33) .. (357.9,215.57) .. controls (358.65,217.81) and (357.9,219.3) .. (355.66,220.04) .. controls (353.42,220.78) and (352.67,222.27) .. (353.42,224.51) .. controls (354.17,226.75) and (353.42,228.24) .. (351.18,228.99) -- (350.53,230.28) -- (350.53,230.28) ;
    \draw    (350.27,90.25) .. controls (351.94,91.91) and (351.94,93.58) .. (350.28,95.25) .. controls (348.62,96.92) and (348.62,98.58) .. (350.29,100.25) .. controls (351.96,101.92) and (351.96,103.58) .. (350.3,105.25) .. controls (348.63,106.92) and (348.63,108.58) .. (350.3,110.25) .. controls (351.97,111.92) and (351.97,113.58) .. (350.31,115.25) .. controls (348.65,116.92) and (348.65,118.58) .. (350.32,120.25) .. controls (351.99,121.92) and (351.99,123.58) .. (350.33,125.25) .. controls (348.67,126.92) and (348.67,128.58) .. (350.34,130.25) .. controls (352.01,131.92) and (352.01,133.58) .. (350.35,135.25) .. controls (348.69,136.92) and (348.69,138.58) .. (350.36,140.25) .. controls (352.03,141.92) and (352.03,143.58) .. (350.37,145.25) .. controls (348.71,146.92) and (348.71,148.58) .. (350.38,150.25) .. controls (352.05,151.92) and (352.05,153.58) .. (350.39,155.25) .. controls (348.73,156.92) and (348.73,158.58) .. (350.4,160.25) .. controls (352.07,161.92) and (352.07,163.58) .. (350.41,165.25) .. controls (348.75,166.92) and (348.75,168.58) .. (350.42,170.25) .. controls (352.09,171.92) and (352.09,173.58) .. (350.43,175.25) .. controls (348.77,176.92) and (348.77,178.58) .. (350.44,180.25) .. controls (352.11,181.92) and (352.11,183.58) .. (350.45,185.25) .. controls (348.79,186.92) and (348.79,188.58) .. (350.46,190.25) .. controls (352.13,191.92) and (352.13,193.58) .. (350.47,195.25) .. controls (348.81,196.92) and (348.81,198.58) .. (350.48,200.25) .. controls (352.15,201.92) and (352.15,203.58) .. (350.49,205.25) .. controls (348.83,206.92) and (348.83,208.58) .. (350.5,210.25) .. controls (352.17,211.92) and (352.17,213.58) .. (350.5,215.25) .. controls (348.84,216.92) and (348.84,218.58) .. (350.51,220.25) .. controls (352.18,221.92) and (352.18,223.58) .. (350.52,225.25) .. controls (348.86,226.92) and (348.86,228.58) .. (350.53,230.25) -- (350.53,230.28) -- (350.53,230.28) ;
    \draw [shift={(350.53,230.28)}, rotate = 89.89] [color={rgb, 255:red, 0; green, 0; blue, 0 }  ][fill={rgb, 255:red, 0; green, 0; blue, 0 }  ][line width=0.75]      (0, 0) circle [x radius= 3.35, y radius= 3.35]   ;
    
    \draw (276.3,75) node [anchor=north west][inner sep=0.75pt]   [align=left] {$\displaystyle \upgamma $};
    \draw (416,75) node [anchor=north west][inner sep=0.75pt]   [align=left] {$\displaystyle \upgamma $};
    \draw (347,75) node [anchor=north west][inner sep=0.75pt]   [align=left] {$\displaystyle \upgamma $};

    \end{tikzpicture}
    \caption{Cubic contact diagram.}
\label{figure:Contact}
\end{figure}

Contact diagrams, see Figure \ref{figure:Contact}, are the easiest to compute. For gravitons they factorise into a polarisation factor which contains the three polarisation tensors for the external gravitons which can be contracted with each other or with spatial momenta, and a \textit{trimmed} wavefunction that carries information about the time evolution. The general form of a cubic contact diagram is therefore \cite{Cabass:2021fnw}\footnote{Here and throughout this work we are dropping the momentum conserving delta function that always appears thanks to the spatial translation symmetry of the theory.}
\begin{align} 
\psi_{3}(\{k \}, \{\bfk \}) = \sum_{\text{contractions}} [e^{h_{1}}(\bfk_{1})e^{h_{2}}(\bfk_{2})e^{h_{3}}(\bfk_{3}) \bfk_{1}^{\alpha_{1}}\bfk_{2}^{\alpha_{2}}\bfk_{3}^{\alpha_{3}}] \psi_{3}^{\text{trimmed}}(\{k \}),
\end{align}
where $h_{i}$ are the helicities of the external fields, we define $\alpha=\alpha_{1}+\alpha_{2}+\alpha_{3}$, and $\{k \}$ and $\{\bfk \}$ collectively denote the external energies and momenta, respectively. Since we are only interested in massless spin-$2$ modes, we have $h_{i} = \pm 2$, and we will concentrate on the $+++$ and $++-$ configurations as the others can be extracted from these by parity transformations \cite{Cabass:2021fnw}. The polarisation factors are easily read off from the Lagrangian and by converting to momentum space, while traditionally the trimmed wavefunction would be computed by integrating the three bulk-boundary propagators, which can be differentiated with respect to time, from the far past at $\eta = - \infty$ to the future boundary at $\eta = 0$. In the far past we Wick rotate such that there is some evolution in Euclidean time which has the effect of projecting onto the vacuum and damping these early time contributions \cite{Maldacena:2002vr}. \\

\noindent Another way of computing $\psi_{3}^{\text{trimmed}}$ was derived in \cite{MLT} which doesn't require any time evolution, which is anyway completely unobservable in the final answer where it has been integrated out, and is referred to as the Manifestly Local Test (MLT). The test requires trimmed wavefunction coefficients to satisfy
 \begin{align}  \label{MLT}
    \frac{\partial }{\partial k_{c}} \psi_{3}^{\text{trimmed}}\Big|_{k_{c}=0}=0\,,\qquad \forall\, c=1,2,3\,, 
    \end{align}
which is simply the statement that the trimmed wavefunction does not contain any terms linear in any of the three external energies (locality also forces there to be no inverse powers of the energies: at cubic order there are no inverse Laplacians to worry about). Note that as we send one of the external energies to zero we do so while holding the others fixed which distinguishes the MLT from soft theorems. The MLT can be used to construct cosmological correlators as was done in \cite{MLT,Cabass:2021fnw,Jain:2021vrv}, but also to verify that the often complicated final results have the correct structure as was done in \cite{Meltzer:2021zin}. It was shown in \cite{Cabass:2021fnw} that general solutions to these equations pick out all possible trimmed wavefunctions, and clearly solving these equations is simpler than having to evaluate the necessary bulk time integrals. The MLT follows from the fact that the bulk-boundary propagator for massless gravitons (and scalars) satisfies
\begin{align}
K_{\gamma}(k,\eta) = (1 - i k \eta) e^{i k \eta}, \qquad 
    \frac{\partial K_{\gamma} }{\partial k} \Big|_{k=0}=0,
\end{align}
and this property is inherited by the trimmed wavefunction since it holds for all $\eta$. The MLT can also be derived by demanding the absence of spurious poles in four-point functions that arise from gluing together two three-point functions. Indeed, such an exchange diagram should be regular as the energy of the exchanged field is taken to zero, and the Cosmological Optical Theorem \cite{COT} implies that this is only the case if the constituent three-point functions satisfy the MLT. We refer the reader to \cite{MLT} for further details on these two derivations.  \\
 
\noindent Now the assumption of a Bunch-Davies vacuum, which we impose throughout, tells us that $\psi_{3}^{\text{trimmed}}$ is a rational function with poles only occurring at $k_{T} = k_{1}+k_{2}+k_{3} = 0$ which is the kinematical limit where energy is conserved. Interestingly, the residue of the leading total-energy pole contains the flat-space scattering amplitude for the same process \cite{Raju:2012zr,Maldacena:2011nz,COT} (in some cases the relationship between correlators and amplitudes is not so straightforward due to enhanced symmetry in the flat-space limit. This is the case for DBI as was shown in \cite{GJS} and confirmed in \cite{InflationaryAdler} at the level of boost-breaking amplitudes.). There can also be logs in the trimmed wavefunction, for low derivative operators, but no other kinematic singularities unless there is some form of non-locality or different vacuum conditions \cite{BBBB}. In this paper we will not encounter logs since in the EFToI there are too many derivatives in the graviton interactions. Indeed the absence of logs in contact diagrams requires $2 n_{\partial_{\eta}} + n_{\partial_{i}} > 3$ \cite{BBBB}, where $n_{\partial_{\eta}}$ and $n_{\partial_{i}}$ are respectively the number of conformal time and space derivatives, and we have seen in Section \ref{GravitonInteractions} that this condition is always satisfied in the EFToI.  \\

\noindent Now to solve the MLT we simply write down an ansatz and organise the solutions in terms of the order of the leading total-energy pole which we denote by $p$ and which counts the number of derivatives in the corresponding vertex \cite{BBBB}. Guidance comes from Bose symmetry: we fix $\psi_{3}^{\text{trimmed}}$ to have the same symmetry as the polarisation part and sum over the remaining permutations once we have constructed $\psi_{3}^{\text{trimmed}}$. This is a consistent thing to do since the MLT is satisfied for all permutations. With the cubic wavefunction coefficient in hand we can compute the corresponding bispectrum $B_{3}$. The relationship is given by \cite{Cabass:2021fnw}\footnote{The COT for cubic contact diagrams is $\psi_{3}(\{k \},\{\bfk \}) + \psi_{3}^{\star}(\{-k \},\{-\bfk \}=0$ \cite{COT} and this dictates which part of the wavefunction contributes to expectation values \cite{Cabass:2021fnw}.}
\begin{align} \label{BispectrumEqn}
B_{3}(\{k \},\{\bfk \}) = - \frac{\psi_{3}(\{k \},\{\bfk \}) + \psi_{3}^{\star}(\{k \},\{-\bfk \})}{ \prod_{a=1}^{3} 2 \text{Re} ~ \psi_{2}(k_{a})},
\end{align}
where $\psi_{2}$ is the quadratic wavefunction coefficient which is perturbatively fixed by the two-derivative quadratic action coming from GR. We have
\begin{align} \label{GRTwoPoint}
\psi_{2,\text{GR}}^{h h'} = \frac{M_{\text{pl}}^2}{H^2} k^3 \delta_{h h'}.
\end{align}
We have dropped imaginary terms in this expression, which are actually divergent at late-times, since they never contribute to correlators. For parity-even interactions, which are the ones of interest here, the numerator is $2 \text{Re} ~ \psi_{3}$ which follows from having an even number of spatial momenta. We can also use $e_{ij}^h(\bfk)^{\ast}= e_{ij}^h(-\bfk)$, which follows from the reality of $\gamma_{ij}(x)$, to see that the polarisation factor is a common factor on both the left and right hand side of this equation. We refer the reader to \cite{Cabass:2021fnw} for more details on deriving this relationship. \\

\noindent We have to compute such contact diagrams for both Type-I and Type-II bispectra. For Type-II we will explain in Section \ref{TypeII} that we can use the MLT to very efficiently write down all allowed wavefunction coefficients by taking into account the fact that each graviton is differentiated with respect to time at least once, while for Type-I, in Section \ref{TypeI} we will directly compute the necessary bulk time integrals. As we can see from \eqref{2BBCT}, the time dependence for two building block operators is quite specific so computing the integrals is more straightforward than finding the necessary subset of MLT solutions. For this reason let us briefly review the Feynman rules for computing contact diagrams.\\

\noindent It is easiest to illustrate the Feynman rules with an example. Consider the graviton interaction
\begin{align}
S_{\text{int}} =  g_{\text{int}} \int d^3 x d \eta ~ a(\eta) \gamma_{ij}' \gamma_{jk}' \gamma_{ki}',
\end{align}
which we take in addition to the free theory coming from GR. This is the leading parity-even correction to the Einstein-Hilbert action in the EFToI. To compute the cubic wavefunction coefficient we convert to momentum space and extract the tensor structure which in this case is simply $e_{ij}(\bfk_{1})e_{jk}(\bfk_{2})e_{ki}(\bfk_{3})$. For each graviton we evolve it from time $\eta$ to the late-time boundary at $\eta_{0}$ by inserting a bulk-boundary propagator $K_{\gamma}$ for each field. We insert an overall $(-i)$ and sum over permutations. This overall factor of $(-i)$ follows from our definition of the wavefunction coefficients in \eqref{WavefunctionOfTheUniverse}. We do this for the two helicity configurations $+++$ and $++-$. For this example, the $+++$ configuration wavefunction coefficient is
\begin{align}
\psi_{3}^{+++} = \frac{6 i g_{\text{int}}}{H} e^{+}_{ij}(\bfk_{1})e^{+}_{jk}(\bfk_{2})e^{+}_{ki}(\bfk_{3})\int^{\eta_{0}}_{-\infty(1-i \eps)} \frac{d \eta}{\eta} ~ K_{\gamma}'(k_{1},\eta)K_{\gamma}'(k_{2},\eta)K_{\gamma}'(k_{3},\eta),
\end{align}
where the minus sign coming from the Feynman rules has been cancelled by the one coming from the scale factor (which appears linearly), and we remind the reader that we don't include the momentum conserving delta function in the wavefunction coefficient c.f. \eqref{WavefunctionOfTheUniverse}. We integrate from the far past to the future boundary and we project onto the vacuum at early times. In the following we will suppress the $i \epsilon$ prescription. Computing the time integral we find
\begin{align}
\psi_{3}^{+++} = -\frac{12 g_{\text{int}}}{H} e^{+}_{ij}(\bfk_{1})e^{+}_{jk}(\bfk_{2})e^{+}_{ki}(\bfk_{3}) \frac{e_{3}^2}{k_{T}^3},
\end{align}
where we have written the fully symmetric time integral in terms of the symmetric polynomials given in \eqref{SymmetricPolys}. This can be easily generalised to any other contact diagram, and as we will explain in Section \ref{SHF}, other helicity configurations can be extracted from the $+++$ one. We nicely see from this example that the degree of the leading order total-energy pole is counting the number of derivatives in the bulk vertex.

\subsection{Single exchange diagrams}
\label{subsec:exchange_diagrams}
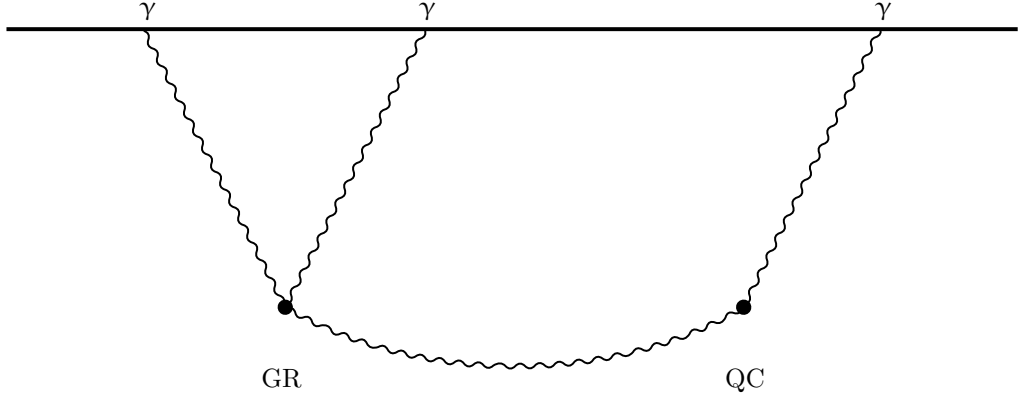
\begin{figure}
    \tikzset{every picture/.style={line width=0.75pt}} 
    
    \begin{tikzpicture}[x=0.75pt,y=0.75pt,yscale=-1,xscale=1]
    uncomment if require: \path (50,280); 
    
    \draw [line width=1.5]    (109,70) -- (619,70) ;
    \draw    (178.26,70.88) .. controls (180.51,71.61) and (181.27,73.09) .. (180.54,75.33) .. controls (179.82,77.57) and (180.58,79.05) .. (182.82,79.78) .. controls (185.06,80.51) and (185.82,81.99) .. (185.09,84.23) .. controls (184.37,86.47) and (185.13,87.95) .. (187.37,88.68) .. controls (189.62,89.41) and (190.38,90.89) .. (189.65,93.14) .. controls (188.92,95.38) and (189.68,96.86) .. (191.92,97.59) .. controls (194.16,98.32) and (194.92,99.8) .. (194.2,102.04) .. controls (193.48,104.28) and (194.24,105.76) .. (196.48,106.49) .. controls (198.72,107.22) and (199.48,108.7) .. (198.75,110.94) .. controls (198.03,113.18) and (198.79,114.66) .. (201.03,115.39) .. controls (203.27,116.12) and (204.03,117.6) .. (203.31,119.84) .. controls (202.58,122.08) and (203.34,123.57) .. (205.58,124.3) .. controls (207.82,125.03) and (208.58,126.51) .. (207.86,128.75) .. controls (207.14,130.99) and (207.9,132.47) .. (210.14,133.2) .. controls (212.38,133.93) and (213.14,135.41) .. (212.42,137.65) .. controls (211.69,139.89) and (212.45,141.37) .. (214.69,142.1) .. controls (216.93,142.83) and (217.69,144.31) .. (216.97,146.55) .. controls (216.24,148.8) and (217,150.28) .. (219.25,151.01) .. controls (221.49,151.74) and (222.25,153.22) .. (221.52,155.46) .. controls (220.8,157.7) and (221.56,159.18) .. (223.8,159.91) .. controls (226.04,160.64) and (226.8,162.12) .. (226.08,164.36) .. controls (225.35,166.6) and (226.11,168.08) .. (228.35,168.81) .. controls (230.59,169.54) and (231.35,171.02) .. (230.63,173.26) .. controls (229.91,175.5) and (230.67,176.98) .. (232.91,177.71) .. controls (235.16,178.44) and (235.92,179.92) .. (235.19,182.17) .. controls (234.46,184.41) and (235.22,185.89) .. (237.46,186.62) .. controls (239.7,187.35) and (240.46,188.83) .. (239.74,191.07) .. controls (239.02,193.31) and (239.78,194.79) .. (242.02,195.52) .. controls (244.26,196.25) and (245.02,197.73) .. (244.29,199.97) .. controls (243.57,202.21) and (244.33,203.69) .. (246.57,204.42) .. controls (248.81,205.15) and (249.57,206.63) .. (248.85,208.87) -- (249.6,210.35) -- (249.6,210.35) ;
    \draw    (320,70.35) .. controls (320.74,72.58) and (319.99,74.07) .. (317.75,74.81) .. controls (315.51,75.55) and (314.76,77.04) .. (315.51,79.28) .. controls (316.25,81.52) and (315.5,83.01) .. (313.26,83.75) .. controls (311.02,84.48) and (310.27,85.97) .. (311.01,88.21) .. controls (311.76,90.45) and (311.01,91.94) .. (308.77,92.68) .. controls (306.53,93.42) and (305.78,94.91) .. (306.52,97.15) .. controls (307.27,99.39) and (306.52,100.88) .. (304.28,101.62) .. controls (302.04,102.35) and (301.29,103.84) .. (302.03,106.08) .. controls (302.77,108.32) and (302.02,109.81) .. (299.78,110.55) .. controls (297.54,111.29) and (296.79,112.78) .. (297.54,115.02) .. controls (298.28,117.26) and (297.53,118.75) .. (295.29,119.48) .. controls (293.05,120.22) and (292.3,121.71) .. (293.04,123.95) .. controls (293.79,126.19) and (293.04,127.68) .. (290.8,128.42) .. controls (288.56,129.16) and (287.81,130.65) .. (288.55,132.89) .. controls (289.29,135.12) and (288.54,136.61) .. (286.31,137.35) .. controls (284.07,138.09) and (283.32,139.58) .. (284.06,141.82) .. controls (284.8,144.06) and (284.05,145.55) .. (281.81,146.29) .. controls (279.58,147.03) and (278.83,148.52) .. (279.57,150.75) .. controls (280.31,152.99) and (279.56,154.48) .. (277.32,155.22) .. controls (275.08,155.96) and (274.33,157.45) .. (275.07,159.69) .. controls (275.81,161.92) and (275.06,163.41) .. (272.83,164.15) .. controls (270.59,164.89) and (269.84,166.38) .. (270.58,168.62) .. controls (271.33,170.86) and (270.58,172.35) .. (268.34,173.09) .. controls (266.1,173.83) and (265.35,175.32) .. (266.09,177.56) .. controls (266.83,179.8) and (266.08,181.29) .. (263.84,182.02) .. controls (261.6,182.76) and (260.85,184.25) .. (261.6,186.49) .. controls (262.34,188.73) and (261.59,190.22) .. (259.35,190.96) .. controls (257.11,191.69) and (256.36,193.18) .. (257.1,195.42) .. controls (257.85,197.66) and (257.1,199.15) .. (254.86,199.89) .. controls (252.62,200.63) and (251.87,202.12) .. (252.61,204.36) .. controls (253.35,206.59) and (252.6,208.08) .. (250.37,208.82) -- (249.6,210.35) -- (249.6,210.35) ;
    \draw [shift={(249.6,210.35)}, rotate = 116.7] [color={rgb, 255:red, 0; green, 0; blue, 0 }  ][fill={rgb, 255:red, 0; green, 0; blue, 0 }  ][line width=0.75]      (0, 0) circle [x radius= 3.35, y radius= 3.35]   ;
    \draw    (249.6,210.35) .. controls (252.45,210.07) and (254.15,211.04) .. (254.7,213.25) .. controls (255.29,215.46) and (256.68,216.2) .. (258.85,215.46) .. controls (261,214.68) and (262.4,215.38) .. (263.07,217.57) .. controls (263.78,219.76) and (265.21,220.44) .. (267.36,219.59) .. controls (270.19,219.04) and (272.01,219.84) .. (272.8,221.99) .. controls (273.64,224.14) and (275.11,224.74) .. (277.22,223.79) .. controls (279.3,222.82) and (280.79,223.39) .. (281.69,225.51) .. controls (282.64,227.62) and (284.15,228.16) .. (286.22,227.12) .. controls (288.27,226.05) and (289.79,226.56) .. (290.79,228.65) .. controls (291.84,230.72) and (293.76,231.31) .. (296.57,230.41) .. controls (298.57,229.25) and (300.13,229.69) .. (301.24,231.72) .. controls (302.39,233.75) and (303.96,234.15) .. (305.95,232.93) .. controls (307.91,231.69) and (309.49,232.06) .. (310.69,234.05) .. controls (311.93,236.03) and (313.52,236.37) .. (315.47,235.08) .. controls (317.38,233.76) and (318.98,234.07) .. (320.27,236) .. controls (321.6,237.93) and (323.21,238.2) .. (325.1,236.83) .. controls (326.97,235.44) and (328.59,235.68) .. (329.96,237.57) .. controls (331.37,239.44) and (332.99,239.66) .. (334.83,238.21) .. controls (336.65,236.74) and (338.28,236.93) .. (339.73,238.76) .. controls (341.21,240.58) and (342.85,240.73) .. (344.64,239.21) .. controls (346.39,237.68) and (348.03,237.8) .. (349.55,239.57) .. controls (351.11,241.33) and (352.75,241.42) .. (354.48,239.83) .. controls (356.99,238.26) and (359.05,238.32) .. (360.65,240.02) .. controls (362.28,241.71) and (363.92,241.73) .. (365.59,240.07) .. controls (367.22,238.4) and (368.86,238.38) .. (370.52,240.02) .. controls (372.21,241.65) and (373.86,241.6) .. (375.45,239.88) .. controls (377.01,238.15) and (378.65,238.07) .. (380.37,239.64) .. controls (382.12,241.2) and (383.76,241.09) .. (385.29,239.31) .. controls (386.78,237.52) and (388.41,237.37) .. (390.19,238.88) .. controls (392,240.37) and (393.62,240.2) .. (395.07,238.36) .. controls (396.48,236.51) and (398.1,236.3) .. (399.93,237.74) .. controls (401.79,239.16) and (403.41,238.92) .. (404.78,237.03) .. controls (406.11,235.12) and (407.72,234.85) .. (409.59,236.22) .. controls (411.5,237.57) and (413.1,237.27) .. (414.38,235.32) .. controls (415.63,233.36) and (417.21,233.03) .. (419.14,234.32) .. controls (421.09,235.59) and (423.06,235.12) .. (425.04,232.93) .. controls (426.19,230.92) and (427.75,230.52) .. (429.72,231.72) .. controls (431.72,232.9) and (433.26,232.46) .. (434.35,230.41) .. controls (435.4,228.36) and (436.93,227.89) .. (438.94,229.01) .. controls (440.98,230.1) and (442.5,229.6) .. (443.49,227.51) .. controls (444.44,225.42) and (445.93,224.89) .. (447.98,225.92) .. controls (450.06,226.92) and (451.91,226.21) .. (453.53,223.79) .. controls (454.36,221.66) and (455.81,221.06) .. (457.9,221.99) .. controls (460.01,222.89) and (461.45,222.25) .. (462.21,220.08) .. controls (462.92,217.92) and (464.34,217.26) .. (466.46,218.09) .. controls (468.61,218.89) and (470.35,218.01) .. (471.68,215.46) .. controls (472.26,213.26) and (473.63,212.52) .. (475.78,213.25) .. controls (477.95,213.93) and (479.3,213.16) .. (479.81,210.94) -- (480.8,210.35) ;
    \draw    (549.6,69.55) .. controls (550.37,71.78) and (549.63,73.27) .. (547.4,74.04) .. controls (545.17,74.81) and (544.44,76.3) .. (545.21,78.53) .. controls (545.97,80.76) and (545.24,82.25) .. (543.01,83.02) .. controls (540.78,83.79) and (540.05,85.29) .. (540.82,87.52) .. controls (541.58,89.75) and (540.85,91.24) .. (538.62,92.01) .. controls (536.39,92.78) and (535.66,94.27) .. (536.43,96.5) .. controls (537.19,98.73) and (536.46,100.22) .. (534.23,100.99) .. controls (532,101.76) and (531.27,103.26) .. (532.04,105.49) .. controls (532.8,107.72) and (532.07,109.21) .. (529.84,109.98) .. controls (527.61,110.75) and (526.88,112.24) .. (527.65,114.47) .. controls (528.41,116.7) and (527.68,118.19) .. (525.45,118.96) .. controls (523.22,119.73) and (522.49,121.23) .. (523.26,123.46) .. controls (524.02,125.69) and (523.29,127.18) .. (521.06,127.95) .. controls (518.83,128.72) and (518.1,130.21) .. (518.87,132.44) .. controls (519.63,134.67) and (518.9,136.16) .. (516.67,136.93) .. controls (514.44,137.7) and (513.71,139.19) .. (514.48,141.42) .. controls (515.25,143.65) and (514.51,145.15) .. (512.28,145.92) .. controls (510.05,146.69) and (509.32,148.18) .. (510.09,150.41) .. controls (510.85,152.64) and (510.12,154.13) .. (507.89,154.9) .. controls (505.66,155.67) and (504.93,157.16) .. (505.7,159.39) .. controls (506.47,161.62) and (505.73,163.12) .. (503.5,163.89) .. controls (501.27,164.66) and (500.54,166.15) .. (501.31,168.38) .. controls (502.07,170.61) and (501.34,172.1) .. (499.11,172.87) .. controls (496.88,173.64) and (496.15,175.13) .. (496.92,177.36) .. controls (497.69,179.59) and (496.95,181.09) .. (494.72,181.86) .. controls (492.49,182.63) and (491.76,184.12) .. (492.53,186.35) .. controls (493.29,188.58) and (492.56,190.07) .. (490.33,190.84) .. controls (488.1,191.61) and (487.37,193.1) .. (488.14,195.33) .. controls (488.91,197.56) and (488.17,199.06) .. (485.94,199.83) .. controls (483.71,200.6) and (482.98,202.09) .. (483.75,204.32) .. controls (484.51,206.55) and (483.78,208.04) .. (481.55,208.81) -- (480.8,210.35) -- (480.8,210.35) ;
    \draw [shift={(480.8,210.35)}, rotate = 116.04] [color={rgb, 255:red, 0; green, 0; blue, 0 }  ][fill={rgb, 255:red, 0; green, 0; blue, 0 }  ][line width=0.75]      (0, 0) circle [x radius= 3.35, y radius= 3.35]   ;
    
    \draw (175,55) node [anchor=north west][inner sep=0.75pt]   [align=left] {$\displaystyle \upgamma $};
    \draw (316,55) node [anchor=north west][inner sep=0.75pt]   [align=left] {$\displaystyle \upgamma $};
    \draw (546,55) node [anchor=north west][inner sep=0.75pt]   [align=left] {$\displaystyle \upgamma $};
    \draw (236.16,240) node [anchor=north west][inner sep=0.75pt]   [align=left] {GR};
    \draw (470,240) node [anchor=north west][inner sep=0.75pt]   [align=left] {QC};

    \end{tikzpicture}
    \caption{Single exchange diagram.}
\label{figure:2BBexchangediagram}
\end{figure}

Since we have corrections to the quadratic action, there are other diagrams in addition to contact ones that can contribute to the cubic wavefunction coefficient at tree-level. In principle there are infinitely many new diagrams which would ultimately resum into a single contact diagram with a new propagator which takes into account the quadratic corrections. However, we are treating these corrections perturbatively in which case only one diagram contributes at leading order. This diagram is shown in Figure \ref{figure:2BBexchangediagram} and corresponds to a single exchange process with a single cubic interaction connected to a quadratic correction (QC) vertex by a bulk-bulk propagator. For two building block operators the largest of these diagrams comes from taking the cubic vertex to be the GR one \cite{Cabass:2021fnw}. Indeed, we are treating the corrections to the quadratic action perturbatively so the largest contribution comes at linear order in these new couplings for which we need to take the cubic vertex to be independent of this small coupling: the GR vertex scales like $1 / P_{\gamma}$, where $P_{\gamma}$ is the power spectrum of the graviton, while the bulk-bulk propagator scales as $P_{\gamma}$, so the product of the two is $\mathcal{O}(1)$. At this order, this exchange diagram is of a comparable size to the contact diagram that comes from these two building block operators, and cancellations are required between these two diagrams for the EFToI consistency relations to be satisfied. This makes sense since the quadratic and cubic operators in the action are tied together by spatial diffeomorphisms which is where these consistency relations come from \cite{Hinterbichler:2013dpa,Maldacena:2002vr,Creminelli:2012ed}. This was checked explicitly for the parity-odd Chern-Simons term in the EFToI in \cite{Cabass:2021fnw}, and we will check it in Section \ref{TypeI} for these parity-even bispectra. \\

\noindent Single exchange diagrams are more complicated to compute compared to contact ones since there are now nested time integrals. Let us illustrate the Feynman rules for computing Figure \ref{figure:2BBexchangediagram} by taking the cubic vertex to be GR, and the quadratic mixing vertex to be one of the terms in \eqref{2BBCTZeroM} where we have $m=0$. As always we introduce an overall factor of $(-i)$, a bulk-boundary propagator for each external line, and a bulk-bulk propagator for the internal line. We act on these propagators with derivatives according to the bulk vertices, add the appropriate polarisation tensors, and sum over permutations. For our example of interest here we have
\begin{align}
\psi^{(n,m=0)}_{3,\text{exchange}} &= -i \times \frac{ M_{\text{pl}}^2}{8}(-2 e_{ik}^{h_{1}}e_{jl}^{h_{2}}k^{3}_{k}k^{3}_{l}e_{ij}^{h_{3}}+e_{ij}^{h_{1}}e_{kl}^{h_{2}}k^{3}_{k}k^{3}_{l}e_{ij}^{h_{3}} + 5 ~ \text{perms})\times 4 \times \frac{M_{\text{pl}}^2 g_{n,0}}{4} \nonumber  \\
&\times \int d \eta d \eta' a^{2}(\eta)a^{2-2n}(\eta') K_{\gamma}(k_{1},\eta)K_{\gamma}(k_{2},\eta) \nonumber  \\ &\times \left[K^{(2n+1)}_{\gamma}(k_{3},\eta') \frac{\partial}{\partial \eta'}G(k_{3},\eta,\eta')+K'_{\gamma}(k_{3},\eta')  \frac{\partial^{2n+1}}{\partial \eta'^{2n+1}} G(k_{3},\eta,\eta')\right] + 2 ~ \text{perms},
\end{align}
where the factor of $4$ in the first line comes from applying \eqref{polnorm} to the quadratic mixing, and the bulk-bulk propagator is given by (see e.g. \cite{COT})
\begin{align}
G(k,\eta,\eta') = 2 P_{\gamma}(k)[\theta(\eta-\eta')K(k, \eta') \text{Im} K(k, \eta) + (\eta \leftrightarrow \eta')],
\end{align}
where $\text{Im} K(k, \eta)$ is the imaginary part of the bulk-boundary propagator which takes the form 
\begin{align}
\text{Im} K(k, \eta) = -\frac{i}{2}[(1 - i k \eta) e^{i k \eta} - (1 + i k \eta) e^{-i k \eta}].
\end{align}
Computing and analysing such time integrals is far simpler when there are no time derivatives on the bulk-bulk propagator. We can guarantee this by integrating by parts and, as always, we can drop all boundary terms. We then have 
\begin{align} \label{ZeromExch}
\psi^{(n,m=0)}_{3,\text{exchange}} = - \frac{i M_{\text{pl}}^4 g_{n,0}}{8}(2 e_{ik}^{h_{1}}e_{jl}^{h_{2}}k^{3}_{k}k^{3}_{l}e_{ij}^{h_{3}}-e_{ij}^{h_{1}}e_{kl}^{h_{2}}k^{3}_{k}k^{3}_{l}e_{ij}^{h_{3}} + 5 ~ \text{perms}) \mathcal{I}_{n,0}(k_{1},k_{2},k_{3}) + 2 ~ \text{perms},
\end{align}
where we have defined
\begin{align} \label{TimeZerom}
\mathcal{I}_{n,0}(k_{1},k_{2},k_{3}) =& \int d \eta d \eta' a^{2}(\eta)K_{\gamma}(k_{1},\eta)K_{\gamma}(k_{2},\eta)G(k_{3}, \eta, \eta') \nonumber \\ & \times \left[\frac{\partial}{\partial \eta'}(a^{2-2n}(\eta')K^{(2n+1)}_{\gamma}(k_{3},\eta') )+\frac{\partial^{2n+1}}{\partial \eta'^{2n+1}}(a^{2-2n}(\eta')K'_{\gamma}(k_{3},\eta') )  \right].
\end{align}
A very similar expression also applies when we take quadratic mixing vertices with $m \neq 0$ from \eqref{2BBCT}, and is given by
\begin{align} \label{NonZeromExch}
\psi^{(n,m \geq 1)}_{3,\text{exchange}} = & \frac{(-1)^{m+1}i M_{\text{pl}}^4 g_{n,m}}{8}(2 e_{ik}^{h_{1}}e_{jl}^{h_{2}}k^{3}_{k}k^{3}_{l}e_{ij}^{h_{3}}-e_{ij}^{h_{1}}e_{kl}^{h_{2}}k^{3}_{k}k^{3}_{l}e_{ij}^{h_{3}} + 5 ~ \text{perms}) k_{3}^{2m} \mathcal{I}_{n,m}(k_{1},k_{2},k_{3}) \nonumber \\~~~~~~& + 2 ~ \text{perms},
\end{align}
where
\begin{align} \label{TimeNonZerom}
\mathcal{I}_{n,m}(k_{1},k_{2},k_{3}) =& \int d \eta d \eta' a^{2}(\eta)K_{\gamma}(k_{1},\eta)K_{\gamma}(k_{2},\eta)G(k_{3}, \eta, \eta') \nonumber \\ & \times \left[\frac{\partial}{\partial \eta'}(a^{q}(\eta')K^{(n+1)}_{\gamma}(k_{3},\eta') )+ (-1)^{n}\frac{\partial^{n+1}}{\partial \eta'^{n+1}}(a^{q}(\eta')K'_{\gamma}(k_{3},\eta') )  \right].
\end{align}
We remind the reader that $q = 2-2m-n \leq 0$. In all cases it is easy to see that every time integral we need to compute for these single exchange diagrams is of the form 
\begin{align} \label{SingleExchangeIntegral}
\mathcal{M}(\alpha,\beta) = \int d \eta d \eta' ~ a^{2}(\eta) K_{\gamma}(k_{1},\eta)K_{\gamma}(k_{2},\eta) G(k_{3},\eta,\eta') a^{\alpha}(\eta') K_{\gamma}^{(\beta)}(k_{3},\eta'),
\end{align}
with $\alpha \leq 0$ and $\beta \geq 1$. We will refer to this integral as the \textit{master time integral}. For $m=0$ we have
\begin{align}
&\mathcal{I}_{n,0}(k_{1},k_{2},k_{3}) = \int d \eta d \eta' a^{2}(\eta)K_{\gamma}(k_{1},\eta)K_{\gamma}(k_{2},\eta)G(k_{3}, \eta, \eta') a^{2-2n}(\eta') \times \nonumber \\ &  \left[K^{(2n+2)}_{\gamma}(k_{3},\eta')+(2-2n)Ha(\eta')K^{(2n+1)}_{\gamma}(k_{3},\eta') + \sum_{k=0}^{2n-2} c_{n,k}H^{k}  a^{k}(\eta')K^{(2+2n-k)}_{\gamma}(k_{3},\eta') \right],
\end{align}
where we have defined
\begin{align}
c_{n,k} = (-1)^k k! {{2n+1}\choose{k}} {{2n-2}\choose{k}},
\end{align}
and from this expression we can write
\begin{align}  \label{Compact1}
\mathcal{I}_{n,0}(k_{1},k_{2},k_{3}) & =   \mathcal{M}(2-2n,2n+2)+(2-2n)H \mathcal{M}(3-2n,2n+1) \nonumber \\
&+\sum_{k=0}^{2n-2}c_{n,k}H^{k}\mathcal{M}(k+2-2n,2+2n-k).
\end{align}
Similarly, for $m \neq 0$ we can write
\begin{align}
 \mathcal{I}_{n,1}(k_{1},k_{2},k_{3}) &=  \mathcal{M}(-n,n+2)  -n H \mathcal{M}(1-n,n+1)+\sum_{k=0}^{n}d_{n,1,k}H^{k}\mathcal{M}(k-n,n+2-k), \label{Compact2} \\
 \mathcal{I}_{n,m}(k_{1},k_{2},k_{3}) &=  \mathcal{M}(2-2m-n,n+2) + (2-2m-n)H \mathcal{M}(3-2m-n,n+1) \nonumber \\
&+\sum_{k=0}^{n+1}d_{n,m,k}H^{k}\mathcal{M}(k+2-2m-n,n+2-k), \qquad m \geq 2, \label{Compact3}
\end{align}
where we have defined
\begin{align}
d_{n,m,k} = (-1)^{n+k} k! {{n+1}\choose{k}} {{n+2m-2}\choose{k}}.
\end{align}
\noindent We will essentially dedicate Section \ref{sec:5.1} to computing $\mathcal{M}(\alpha,\beta)$ and then finding the final form of these single exchange diagrams, but let us first derive some general properties. We will show that $\mathcal{M}(\alpha,\beta)$ is only singular at $k_{T} = 0$, and satisfies the MLT for each external leg. These two properties are enough for us to conclude that single exchange diagrams in the EFToI have the same structure as cubic contact diagrams and are therefore captured by the general analysis of \cite{Cabass:2021fnw}. It would be interesting to investigate if this holds for other cubic diagrams too i.e. those with more bulk-bulk propagators.   \\

\noindent To study this master time integral we need expressions for the derivatives of the bulk-boundary propagator. We have
\begin{align} \label{Kderivs}
K^{(\beta)}(k,\eta) = \frac{\partial^\beta}{\partial \eta ^\beta} K(k,\eta) = (ik)^\beta (1-\beta - i k \eta) e^{i k \eta}.
\end{align}
Let's first ask what singularities $\mathcal{M}(\alpha,\beta)$ can have. As has been discussed in the literature in several places, see e.g. \cite{CosmoBootstrap3}, exchange diagrams at tree-level have a restricted set of singularities: they can be singular when the energy of all external legs sums to zero, and when the energy of an individual vertex sums to zero. This is a consequence of having local vertices and Bunch-Davies initial conditions. For a diagram like ours, in general the allowed singularities are therefore at $k_{T} = 0$ and $k_{3} = 0$. How can $k_{3} = 0$ singularities arise? If we first perform the $\eta$ integral, meaning that we take the $\theta(\eta'-\eta)$ part of the bulk-bulk propagator, then no $k_{3}=0$ singularities can arise. Indeed, we would integrate from the far past at $\eta = - \infty$ up to $\eta'$ and in the process singularities can only arise from exponential factors, and their arguments will always contain a sum of energies. The subsequent $\eta'$ integral also cannot yield $k_{3}=0$ poles for the same reason. If we do the $\eta'$ integral first, however, then it does look like inverse powers of $k_{3}$ are possible. The relevant part of the integral is 
\begin{align}
\frac{1}{k_{3}^3}\int_{-\infty}^{\eta} d \eta' ~  k_{3}^{\beta} \eta'^{-\alpha}(1 - i k_{3}\eta')(1-\beta - i k_{3}\eta')e^{2 i k_{3} \eta'},
\end{align}
where the factor of $k_{3}^{-3}$ comes from the power spectrum in the bulk-bulk propagator. This integral contributes various powers of $k_{3}$ with the smallest power given by $(\beta + \alpha-4)$. For $m=0$ we have $\beta + \alpha-4=0$ so there are no inverse powers of $k_{3}$, while for $m \neq 0$ we have $\beta +\alpha-4=-2m$. However, the spatial derivatives in this case introduce an additional factor of $k_{3}^{2m}$ so again there are no inverse powers of $k_{3}$. No further inverse powers of $k_{3}$ can be generated when we come to now do the $\eta$ integral since now all exponents contain sums of energies. Any folded singularities drop out when we compute the full integral, so only total-energy poles are allowed. We therefore conclude that in the EFToI there are too many derivatives for singularities to occur as $k_{3} \rightarrow 0$. \\

\noindent What about logs? We showed in the previous section that contact diagrams in the EFToI can never have logs since there are too many derivatives in the corresponding vertices. This analysis does not apply to single exchange diagrams, but again we don't expect logs to arise from the corresponding time integrals. If such a log did arise from one of these single exchange diagrams, it could not be cancelled by a contact diagram and since we are computing parity-even correlators, it would contribute to the bispectra. This was proven in generality in \cite{Cabass:2021fnw} where it was shown that only for parity-even interactions can bispectra have a log. Furthermore, the log would also need to depend on a sum of energies. Indeed, if it was to just depend on $k_{3}$ it would need to come from first integrating with respect to $\eta'$ but because we have at least four derivatives, and therefore non-negative powers of $\eta'$, that part of the integral cannot give rise to a log. The presence of a log would be felt in the leading soft theorem which is given by
\begin{align}
\langle \gamma^{h_{s}}_{\bfk - \bfq / 2} \gamma^{h_{s}}_{-\bfk - \bfq / 2} \gamma^{h_{l}}_{\bfq} \rangle \sim \frac{3}{2} P^{h_{l}}(q) P^{h_{s}}(k) e_{ij}^{h_{l}} \hat{k}_{i} \hat{k}_{j},
\end{align}
for some soft momentum $\bfq$. The right hand side of this soft theorem does not have a log dependence so there also cannot be a log on the left\footnote{A similar argument can be used to show that there cannot be any singularities at $k_{3} = 0$.}. This shows that the cubic wavefunction coefficient, and therefore single exchange diagrams in the EFToI, cannot have log singularities. We note that the absence of logs might be highly non-trivial, and require cancellations between different parts of the full integral in \eqref{SingleExchangeIntegral}. As an example, consider the case where $m=2$ and $n=0$ i.e. the quadratic correction only has spatial derivatives. If we first compute the $\theta(\eta - \eta')$ part of the integral then we find that at late times there is a log dependence of the form $\log[(k_{T}+2k_{3}) / k_{T}]$, while the $\theta(\eta' - \eta)$ part of the integral contributes an equal and opposite term such that no logs appear in the final result. \\

\noindent What about the MLT? It is easy to see that the MLT is satisfied for $k_{1}$ and $k_{2}$ since they only appear through the bulk-boundary propagators which as we explained above ensures that the MLT is satisfied for those energies. Now for $k_{3}$, the only dependence that we didn't take into account above was from $\text{Im} K(k_{3}, \eta)$ which when expanded for small $k_{3}$ starts at $\mathcal{O}(k_{3}^3)$. It follows that the result of this integral (multiplied by any factors coming from spatial derivatives) always contains a factor of $k_{3}^3$, from both theta functions, which in turn ensures that it satisfies the MLT for this leg too. So we conclude that the leading single exchange diagrams in the EFToI only have poles as $k_{T} \rightarrow 0$ and they always satisfy the MLT for all three legs. It follows that the final result of these diagrams is captured by the analysis of \cite{Cabass:2021fnw}.\\

\noindent Finally, let us outline how one goes from such an exchange contribution to the wavefunction to expectation values. This was worked out in detail in \cite{Cabass:2021fnw} where it was shown that the correction to the two-point function yields a slightly more complicated relation compared to the case for contact diagrams. Up to linear order in the correction to the two-point function,  the relationship is 
\begin{align} \label{WFtoCorrelator}
B_{3} = \frac{1}{\Pi_{i=1}^3 \mathcal{P}_{2}^{(0)}(k_{i})} \left(- \mathcal{P}_{3}^{\{ \lambda_{i} \}}( \{ \bfk \}, \{ k \}) + \mathcal{P}_{3}^{\{ \lambda_{i} \}}( \{ \bfk \}, \{ k \}) \left(\frac{\delta \mathcal{P}_{2}^{\lambda_{1}}(k_{1})}{\mathcal{P}_{2}^{(0)}(k_{1})} + 2 ~ \text{perms} \right) \right),
\end{align}
where the permutations are of both momenta and helicity labels, and we have defined the combination
\begin{align}
\mathcal{P}_{n}^{ \{ \mu_{i} \}}( \{ \bfk \}, \{ k \}) = \psi_{n}^{ \{ \mu_{i} \}}( \{ \bfk \}, \{ k \}) + \psi_{n}^{ \{ \mu_{i} \}}( \{ -\bfk \}, \{ k \})^{\star}.
\end{align}
Here $\mathcal{P}_{2}^{(0)}$ is the GR contribution, with the two-point function given by \eqref{GRTwoPoint}, while $\delta \mathcal{P}_{2}$ comes from the small quadratic corrections. Note that this formula includes all tree-level contributions to $\psi_{3}$, so both contact and exchange diagrams are added. In Section \ref{TypeI} we will use this formula to convert our calculations of wavefunction coefficients into bispectra.

\subsection{Boundary terms} \label{BoundaryTerms}

Before computing the bispectra, let us first verify that boundary terms evaluated at $\eta=0$ cannot contribute to the late-time wavefunction. Such terms could arise either independently, or as a consequence of the integration by parts we have performed to arrive at \eqref{TwoBB}. The general forms of boundary contributions from two and three building block boundary operators are, schematically,
\be \label{eq:2-3BB-BoundaryTerms}
\lim\limits_{\eta \to 0^-} \left( \eta^{a+b-3} \O^{(a)} A \cdot \O^{(b)} B \right) , \quad  \lim\limits_{\eta \to 0^-} \left( \eta^{a+b+c-3} \O^{(a)} A \cdot \O^{(b)} B \cdot  \O^{(c)} C \right)\,,
\ee
where $a,b,c$ denote the number of derivatives in the corresponding operators, $A,B,C \in \{\delta K^i_{\ j}, \eta^2 \tilde{R}^i_{\ j}\}$ are scale-invariant objects, and indices in \eqref{eq:2-3BB-BoundaryTerms} have been suppressed. The powers of $\eta$ are fixed by scale invariance. In conformal time, the extrinsic curvature is given by
\be
\delta K^{i}_{\ j} = -\frac{1}{2} a^2 g^{il} H \eta \gamma'_{mn} \frac{\delta \left(   e^{\gamma}  \right)_{lj}  }{\delta \gamma_{mn}} \,,
\ee
and given that the graviton equation of motion dictates that asymptotically $\gamma'_{ij} \sim \mathcal{O}(\eta)$ as $\eta \to 0^-$, we have
\be
\delta K^{i}_{\ j} \sim \mathcal{O}(\eta^2) \quad \text{as} \quad  \eta \to 0^- \,.
\ee
Here we have used the fact that $a^2 g^{il}$ is constant at late-times. Similarly, for the scale-invariant Ricci tensor we have
\be
\eta^2 \tilde{R}^{i}_{\ j} \sim \mathcal{O}(\eta^2) \quad \text{as} \quad  \eta \to 0^- \,.
\ee
This late-time behaviour of the building blocks ensures that the boundary terms in \eqref{eq:2-3BB-BoundaryTerms} approach zero at the future boundary. Indeed, close to $\eta = 0$ we have
\begin{eqnarray} \nn
\eta^{a+b-3} \prod\limits_{a,b} \O^{(x)} A& \sim & \O\left( \eta^{a+b-3 +  \sum_{a,b} \max\{ 2-x, 0 \}} \right) \sim \O\left( \eta^{1 +\sum_{a,b} \max\{0, x-2 \}} \right) \sim \O(\eta), \\ \nn
\eta^{a+b+c-3} \prod\limits_{a,b,c} \O^{(x)} A & \sim & \O\left( \eta^{a+b+c-3 +  \sum_{a,b,c} \max\{ 2-x, 0 \}} \right) \sim \O\left( \eta^{3 +\sum_{a,b,c} \max\{0, x-2 \}} \right) \sim \O(\eta^3)\,.
\end{eqnarray}
This shows that we can freely integrate by parts at the level of these covariant building blocks, as we have done in Section \ref{EFToI}, without having to worry about boundary contributions to the wavefunction. \\

\noindent A question distinct from the one we have just discussed, is whether we could have generated any boundary terms when we integrated by parts in Section \ref{subsec:exchange_diagrams} to remove all time derivatives from the bulk-bulk propagator to the bulk-boundary one, c.f. \eqref{SingleExchangeIntegral}. For clarity, recall that the nested time integral of interest is
\be
\int\limits_{-\infty}^0 d \eta' \alpha(\{ k_i \}, \eta')  \int\limits_{-\infty}^0 d \eta ~ k^{n}  G^{(a)}(k; \eta, \eta') K^{(b)}(k; \eta) \eta^{n+a+b-4}\,,
\ee
where $a, b \geqslant 1$, because $\delta K_{ij} \propto \gamma'_{ij}$ at linear order so each propagator must include at least one time derivative. Again the power of $\eta$ is fixed by scale invariance, and the factor of $k^{n}$ originates from $n$ spatial derivatives acting on either line attached to the quadratic vertex. The $\eta'$ vertex is the one coming from GR which won't affect the below discussion since it doesn't contain any time derivatives. Our goal is to use integration by parts to eliminate all conformal time derivatives from $G$. We have
\begin{eqnarray} \nn
\int\limits_{-\infty}^0 d \eta ~ G^{(a)}(\eta) K^{(b)}(\eta) \eta^{n_s+a+b-4} = 
(-1)^a \int\limits_{-\infty}^0 d \eta ~ G(\eta) \partial^a_{\eta} \left( K^{(b)}(\eta) \eta^{n + a + b - 4} \right) \\
+ \sum\limits_{j=0}^{j=a-1} (-1)^j \left[  G^{(a-1-j)}(\eta) \partial^{j}_{\eta} \left( K^{(b)}(\eta) \eta^{n + a + b - 4}     \right) \right]_{\eta \to 0} .
\end{eqnarray}
What are the necessary and sufficient conditions for the boundary terms in the second line to vanish? A useful limit is that as $\eta \to 0$, the bulk-bulk propagator decays as $\eta^3$:
\be
G(k; \eta, \eta') \sim \frac{1}{2} K(k; \eta') \eta^3 + \eta^3 \cdot \mathcal{O}(k \eta) \quad \text{as} \quad \eta \to 0^-.
\ee
First consider the case $a-1-j \leqslant 3$. We have
\begin{eqnarray}
G^{(a-1-j)}(k; \eta, \eta') & \sim & \O \left( \eta^{4 + j - a} \right), \\
\partial^{j}_{\eta} \left( K^{(b)}(k; \eta) \eta^{n_s + a + b - 4}  \right) & = & \mathcal{O}(\eta^{n_s + a + b - 4 - j}),
\end{eqnarray}
so we conclude that
\be
G^{(a-1-j)}(k; \eta, \eta') \partial^{j}_{\eta} \left( K^{(b)}(k; \eta) \eta^{n + a+b - 4}  \right)  =
 \mathcal{O}(\eta^{n + b}).
\ee
Thus, the boundary term decays at least as fast as $\eta^{n + b}$. Since $b \geqslant 1$ (which must hold for operators constructed out of $\delta K^i_{\ j}$), this is sufficient for the boundary term to vanish. \\

\noindent Now consider $a - 1 -j > 3$, which implies $a > 4+j \geqslant 4$. Then for the boundary term to vanish it is necessary and sufficient that
\be
\lim\limits_{\eta \to 0} K^{(b)}(k;\eta) \eta^{n + a + b - 4} = 0 .
\ee
Since $a \geq 4$, the above relation holds for any $b$. \\

\noindent In conclusion, for quadratic operators that appear in the EFToI we can proceed as we did in Section \ref{subsec:exchange_diagrams} and move all time derivatives in the bulk integrals onto the bulk-boundary propagators, which significantly simplifies the calculation we will perform in Section \ref{TypeI}.

\subsection{Spinor-helicity formalism} \label{SHF}

\noindent Once we have the final form of the various time integrals, the last step is to account for the tensor structure. For the single exchange diagram of interest in this paper, the nontrivial tensor structure comes from the GR vertex, while for contact diagrams this structure originates from the EFToI cubic interactions beyond GR. The computation and the final result are much simplified if we work with the de Sitter spinor-helicity formalism developed in \cite{Maldacena:2011nz}. Here we discuss the subject only very briefly and refer the reader to \cite{Maldacena:2011nz,Cabass:2021fnw,CosmoBootstrap3,Dreiner} for further details, with many useful relations given in \cite{CosmoBootstrap3}. \\

\noindent A null four-vector $k_{\mu}$ can be represented by a pair of two-component, non-Grassmanian spinors $(\lambda, \tilde{\lambda})$ as follows:
\be
k_{\alpha \dot{\alpha}} = \sigma^{\mu}_{\alpha \dot{\alpha}} k_{\mu} = \lambda_{\alpha} \tilde{\lambda}_{\dot{\alpha}}\,,
\ee
where we defined $k_{\mu} = (k, \bfk)$ and $\sigma^{\mu} =  (\mathds{1},\bm{\sigma})$ are the Pauli matrices. Here we have constructed a null four-vector from the spatial momenta with $k  = |\bfk|$. In contrast to flat-space, here energy is not conserved so a number of spinor helicity identities in flat-space are altered in de Sitter. For example, conservation of spatial momenta for three particles yields
\begin{align}
\langle ab \rangle [ab] &= k_{T}I_{c}, \qquad \text{for} \quad a \neq b \neq c, \\
\sum_{a=1}^{3} \lambda^{(a)}_{\alpha} \tilde{\lambda}^{(a)}_{\dot{\alpha}} &= k_{T} (\sigma^{0})_{\alpha \dot{\alpha}},
\end{align}
where we have defined the spinor brackets 
\begin{eqnarray}
\langle ab \rangle & = & \epsilon^{\alpha \beta} \lambda^{(a)}_{\alpha} \lambda^{(b)}_{\beta}, \\
{} [ ab ] & = & \epsilon^{\dot{\alpha} \dot{\beta}} \tilde{\lambda}^{(a)}_{\dot{\alpha}} \tilde{\lambda}^{(b)}_{\dot{\beta}}\,,
\end{eqnarray}
and $I_{a} = k_{b}+k_{c}-k_{a}$. The object $(\bar{\sigma}^{0})^{\dot{\alpha}\alpha}$ can be used to pick out of the time component of a vector e.g. $(\bar{\sigma}^{0})^{\dot{\alpha}\alpha}k_{\alpha \dot{\alpha}} = 2 k$. This would not be allowed in a Lorentz invariant theory but is perfectly fine in our cosmological setting. This allows us to define polarisation vectors which have a vanishing time component. If we write the polarization tensors for $\gamma_{ij}$ as $e^{\pm}_{ij}(\bfk) =e^{\pm}_{i}(\bfk) e^{\pm}_{j}(\bfk)$, then we have
\begin{eqnarray}
e^+_{\alpha \dot{\alpha}}(\bfk) & = &  \frac{\left( \sigma^0 \right)_{\alpha \dot{\beta}} \tilde{\lambda}^{\dot{\beta}} \tilde{\lambda}_{\dot{\alpha}}}{ k}\,, \\
e^-_{\alpha \dot{\alpha}}(\bfk) & = & \frac{\left( \sigma^0 \right)_{\beta \dot{\alpha}} \lambda^{\beta} \lambda_{\alpha} } { k}\,.
\end{eqnarray}
The numerical factors in these expressions follow from $e^{h}_{ij}(\bfk) e^{h'}_{ij}(\bfk)^{\ast}=4\delta_{hh'}$, and the simplicity of these expressions motivates this particular normalisation. One can check that these polarisation tensors are transverse to the momentum and have a vanishing time component. For parity-even interactions at three-points, useful formulae for going from polarisations to spinors are 
\begin{align}
e^{a+} \cdot e^{b+} = -\frac{[ab]^2}{2k_{a}k_{b}}, \qquad e^{a-} \cdot e^{b-} = -\frac{\langle ab \rangle^2}{2k_{a}k_{b}}, \qquad e^{a+} \cdot e^{b-} = \frac{I_{b}^2}{2k_{a}k_{b}}\frac{\langle cb \rangle^2}{\langle ca \rangle^2}=\frac{I_{a}^2}{2k_{a}k_{b}}\frac{[ca]^2}{[cb]^2},
\end{align}
\begin{align}
k^{a} \cdot e^{b+} = \frac{I_{b}}{2 k_{b}} \frac{[ab][bc]}{[ac]}, \qquad k^{a} \cdot e^{b-} = \frac{I_{b}}{2 k_{b}} \frac{\langle ab \rangle \langle bc \rangle }{\langle ac \rangle}.
\end{align}
The wavefunction coefficients can then always be expressed in terms of coupling constants, the square and angle brackets, and the energies $k_a$, with the linear combinations $I_{a}$ playing a special role. Let us therefore convert all tensor structures into bracket expressions. The tensor structure part of $\psi_{3, +++}$ always takes the following form \cite{Cabass:2021fnw}
\be
\left( e^+(\bfk_1) e^+(\bfk_2) e^+(\bfk_3) \bfk_1^{\alpha_1} \bfk_2^{\alpha_2} \bfk_3^{\alpha_3} \right) = \frac{[12]^2[23]^2[31]^2}{k_1^2 k_2^2 k_3^2} h_{\alpha}(k_a) \equiv {\rm SH}_{+++} h_{\alpha}(k_a)   \,,
\ee
where $h_{\alpha}$ is a polynomial in the energies, of dimension $\alpha = \alpha_1+\alpha_2+\alpha_3$. For parity even interactions, which we are considering here, possible values are $\alpha = 0,2,4,6$. For each interaction, $h_{\alpha}$ must be found by an explicit calculation. For example, the GR vertex can be represented by $h_{\alpha=2}$. The simplicity of the method lies in the fact that once the $+++$ computation is performed, all other helicity configurations come almost automatically. The time integral part of the wavefunction coefficient is the same as for the $+++$, while the tensor structure part can be found as follows:
\begin{eqnarray}
\left( e^+(\bfk_1) e^+(\bfk_2) e^-(\bfk_3) \bfk_1^{\alpha_1} \bfk_2^{\alpha_2} \bfk_3^{\alpha_3} \right) & = & \frac{[12]^6}{[23]^2 [31]^2} \frac{I_1^2 I_2^2 }{k_1^2 k_2^2 k_3^2} h_{\alpha}(k_1,k_2,-k_3)  \equiv {\rm SH}_{++-} h_{\alpha}(k_1,k_2,-k_3) \nonumber    \,, \\
\left( e^-(\bfk_1) e^-(\bfk_2) e^+(\bfk_3) \bfk_1^{\alpha_1} \bfk_2^{\alpha_2} \bfk_3^{\alpha_3} \right) & = & \frac{\langle 12 \rangle^6}{\langle 23 \rangle^2 \langle 31 \rangle^2} \frac{I_1^2 I_2^2}{k_1^2 k_2^2 k_3^2} h_{\alpha}(k_1,k_2,-k_3)  \equiv {\rm SH}_{--+} h_{\alpha}(k_1,k_2,-k_3)   \,, \nonumber \\
\left( e^-(\bfk_1) e^-(\bfk_2) e^-(\bfk_3) \bfk_1^{\alpha_1} \bfk_2^{\alpha_2} \bfk_3^{\alpha_3} \right) & = & \frac{\langle 12 \rangle^2 \langle 23 \rangle^2 \langle 31 \rangle^2}{k_1^2 k_2^2 k_3^2} h_{\alpha}(k_1,k_2,k_3)  \equiv {\rm SH}_{---} h_{\alpha}(k_a)\,,
\end{eqnarray}
We refer the reader to \cite{Cabass:2021fnw} for a derivation and a more in-depth discussion of this construction. We will present the final form of graviton bispectra using these spinor variables.


\section{Type-I bispectra} \label{TypeI}

We now start computing the graviton bispectra in the EFToI. We begin with the Type-I bispectra that arise from two building block operators. In this case both single exchange and contact diagrams contribute and we deal with them in turn. We then write down the final form of the bispectra which requires us to convert our wavefunction expressions into expectation values. Finally we check that our results satisfy the consistency relations which relate the soft limit of a three-point function to the two-point function\cite{Hinterbichler:2013dpa,Maldacena:2002vr,Creminelli:2012ed}, which now receives corrections. This provides a non-trivial check of our final result.  

\subsection{Single exchange diagram} \label{sec:5.1}
We begin with the exchange contributions to the wavefunction which arise due to the quadratic corrections to the free theory in \eqref{2BBCT}. We have computed these contributions when we gave examples of the Feynman rules for exchange diagrams, and they are given by \eqref{ZeromExch} and \eqref{NonZeromExch}. If we write the polarisation tensors in such a way that we have full symmetry in particles $1$ and $2$, then for the $+++$ configuration we have
\begin{align}
4e_{ik}^{h_{1}}e_{jl}^{h_{2}}k_{k}^{3}k_{l}^{3}e_{ij}^{h_{3}} - e_{ij}^{h_{1}}e_{kl}^{h_{2}}k_{k}^{3}k_{l}^{3}e_{ij}^{h_{3}}-e_{ij}^{h_{2}}e_{kl}^{h_{1}}k_{k}^{3}k_{l}^{3}e_{ij}^{h_{3}} = -\frac{1}{16}\text{SH}_{+++} \left(4I_{1}I_{2} + I_{1}^2+I_{2}^2 \right).
\end{align}
Using this expression in \eqref{ZeromExch} and \eqref{NonZeromExch}, for the $+++$ configuration, we find
\begin{align} \label{ExchSHPPP}
\psi^{+++}_{\text{2BB, exchange}} &= \sum_{n=1} \frac{i M_{\text{pl}}^4 g_{n,0}}{64} \text{SH}_{+++}k_{T}^2 \mathcal{I}_{n,0}(k_{1},k_{2},k_{3}) + 2 ~ \text{perms},  \\
&+ \sum_{n=0,m=1}  \frac{(-1)^{m}i M_{\text{pl}}^4 g_{n,m}}{64} \text{SH}_{+++}k_{T}^2 k_{3}^{2m} \mathcal{I}_{n,m}(k_{1},k_{2},k_{3}) + 2 ~ \text{perms}.
\end{align}
Notice how once we summed over the remaining two permutations in the tensor structure, the contribution reduced simply to $k_{T}^2$. Recall that here we are using the tensor structure of GR where this factor of $k_{T}^2$ is familiar \cite{Maldacena:2011nz} and cancels the $1/k_{T}^2$ that comes from the time integral when computed in pure gravity. This means that the $+++$ configuration for pure gravity in de Sitter space does not have a total-energy pole which can be traced back to the fact that the corresponding amplitude for this configuration is zero. We will see that this behaviour is unique to pure gravity: the EFToI corrections to the bispectrum do indeed have total-energy poles for the $+++$ configuration. As we explained above we can now easily extract the $++-$ configuration which is given by
\begin{align} \label{ExchSHPPM}
\psi^{++-}_{\text{2BB, exchange}} &= \sum_{n=1} \frac{i M_{\text{pl}}^4 g_{n,0}}{64} \text{SH}_{++-} I_{3}^2 \left(  \mathcal{I}_{n,0}(k_{1},k_{2},k_{3}) + 2 ~ \text{perms} \right),  \\
&+ \sum_{n=0,m=1}  \frac{(-1)^{m}i M_{\text{pl}}^4 g_{n,m}}{64} \text{SH}_{++-}I_{3}^2  \left(k_{3}^{2m} \mathcal{I}_{n,m}(k_{1},k_{2},k_{3}) + 2 ~ \text{perms} \right),
\end{align}
where we have used the fact that $k_{T}^2 \rightarrow I_{3}^2$ as we send $k_{3} \rightarrow - k_{3}$. We remind the reader that $\mathcal{I}_{n,0}$ and $\mathcal{I}_{n,m}$ are defined in \eqref{TimeZerom} and \eqref{TimeNonZerom}, which can be compactly written as \eqref{Compact1},\eqref{Compact2} and \eqref{Compact3} in terms of the master time integral \eqref{SingleExchangeIntegral}. Our focus now is on computing this master time integral.\\

\noindent To compute this integral we use the formalism developed in \cite{Meltzer:2021zin} to write down a dispersion formula in terms of discontinuities of the bulk-boundary propagator. In order to ensure a Bunch-Davies vacuum in the infinite past requires the use of the $i\epsilon$ prescription, $k \to k-i\epsilon$, where the norm of $k$ is given a negative imaginary part, i.e. $-k \to -k+i\epsilon$. In polar coordinates, $k^2 = e^{i\theta}$, this becomes the condition $\theta \in (-2\pi,0)$. If $\theta$ is in this interval, then the Feynman integrals converge. It is therefore natural to place the $k^2$ branch cut on the positive real axis. We then define the following monodromy operation
\begin{equation}
     {\rm disc}_{p^2} f(k^2)=f((e^{-i\pi}k)^2)-f(k^2)=f((-k+i\epsilon)^2)-f((k-i\epsilon)^2)=f((-k^*)^2)-f(k^2),
\label{general_discontinuity}
\end{equation}
i.e. the argument of the term on the left hand side comes from rotating $k$ in the complex plane by $\theta=-\pi$. By utilising the Hermitian analyticity of the bulk-boundary propagator, we can express the discontinuity of the bulk-boundary propagator as\footnote{A function is described to be Hermitian analytic provided it satisfies the relation $f^*(-k^*)=f(k)$. We refer the reader to \cite{Goodhew:2021oqg,COT} for further discussion on this topic.}
\begin{equation}
     {\rm disc}_{p^2} K(p,\eta)=K(e^{-i\pi }p,\eta)-K(p,\eta)=K^*(p,\eta)-K(p,\eta)=-2i\,{\rm Im} K(p,\eta).
\label{bulk-to-boundary_discontinuity}
\end{equation}
\noindent We can write a general bulk-bulk propagator $G_{\nu}(k,\eta,\eta')$ in terms of a dispersion formula\cite{Meltzer:2021zin}
\begin{equation}
    G_{\nu}(k,\eta,\eta') = \frac{1}{2\pi i}\int_0^{+\infty} \frac{d p^2}{p^2-k^2+i\epsilon} {\rm disc}_{p^2} G_{\nu}(p,\eta,\eta'),
\label{general_bulk-bulk_dispersion}
\end{equation}
where $\nu$ is the usual order of the Hankel function related to the mass of the bulk field by
\begin{equation}
    \nu = \sqrt{\frac{9}{4}-\frac{m^2}{H^2}} \, .
\label{order_of_Hankel}
\end{equation}
The discontinuity of the bulk-bulk propagator can then be expressed in terms of the discontinuities of the bulk-boundary propagator as
\begin{equation}
     {\rm disc}_{p^2} G_{\nu}(p,\eta,\eta') = iP_{\nu}(p){\rm disc}_{p^2} K(p,\eta){\rm disc}_{p^2} K(p,\eta')\,,
\label{general_bulk-to-bulk_discontinuity}
\end{equation}
where $P_{\nu}$ is the power spectrum.
Given that we are working with massless gravitons in the EFToI, we can take $\nu=3/2$, allowing us to rewrite \eqref{SingleExchangeIntegral} as 
\begin{align}
\label{Meltzer_exchange_diagram}
    \mathcal{M}(\alpha,\beta)=\frac{1}{2\pi i}&\int_{-\infty}^{+\infty} \frac{d p}{p^2-k_3^2+i\epsilon} \frac{i H^2}{2M_{\text{pl}}^2 \, p^2}\int_{-\infty}^{0} d\eta \, (-H\eta)^{-2} K(k_1,\eta) K(k_2,\eta)[K^*(p,\eta)-K(p,\eta)] \nonumber \\ 
    &\int_{-\infty}^{0} d\eta' \, (-H\eta')^{-\alpha} K^{(\beta)}(k_3,\eta') [K^*(p,\eta')-K(p,\eta')]\,.
\end{align}
Two comments are in order. First, we have switched the order of the integrals. This is typically not allowed: the behavior of ${\rm disc}_{p^2} K(p,\eta)$ in the infinite past is such that the $\eta$ integral converges only for particular values of $p^2$. One can carry out the integral for these values and then analytically continue: the final result is that the nested integral becomes an integral in $p^2$ of the product of the discontinuities of single integrals. The second observation is that the product of these discontinuities is an even function of $p$. We can then change variables from $p^2$ to $p$, and take advantage of the fact that the whole integrand is even in $p$ to extend the range of integration from $-\infty$ to $+\infty$ \cite{Meltzer:2021zin}. Carrying out the integrals in $d\eta$ and $d\eta'$ we find 
\begin{equation}
\label{Meltzer_exchange_diagram2}
    \mathcal{M}(\alpha,\beta)=\frac{1}{2\pi i}\int_{-\infty}^{+\infty} d p \; {\cal N}(k_1,k_2,k_3,p) \, ,
\end{equation}
where we have dropped the $i\epsilon$ for simplicity of notation. The function $\cal N$ is given by 
\begin{equation}
\label{Meltzer_exchange_diagram_integrand}
\begin{split}
    {\cal N}(k_1,k_2,k_3,p) = &{-\frac{ iH^{-\alpha}}{M_{\text{pl}}^2}}\frac{p(k_1^2+4 k_1 k_2+k_2^2-p^2)}{(p^2-k_3^2)(k_1+k_2-p)^2 (k_1+k_2+p)^2} \\ 
    &\;\;\;\; \times (i k_3)^{\beta}(-\alpha)!i^\alpha \bigg[\frac{\left((k_3-p) (k_3+p)^{\alpha }-(k_3+p) (k_3-p)^\alpha\right)(\beta-1)}{k^2_3-p^2}\\
    &\hphantom{\;\;\;\;\times (i k_3)^{\beta}(-\alpha)!i^\alpha \bigg[ } + ({-\alpha}+1)\big((k_3-p)^{\alpha-2}(k_3+\beta p-p) - (k_3+p)^{\alpha-2}(k_3-\beta p+p)\big) \\
    &\hphantom{\;\;\;\;\times (i k_3)^{\beta}(-\alpha)!i^\alpha \bigg[ } - 
    ({-\alpha}+2)({-\alpha}+1)k_3p\big((k_3-p)^{\alpha-3}+(k_3+p)^{\alpha-3}\big)\bigg] \,,
    \end{split}
\end{equation}
which is manifestly symmetric under $p\to{-p}$. 
We also see that the integrand in \eqref{Meltzer_exchange_diagram_integrand} vanishes as $p \to \infty$ in the complex plane, and therefore we can close the $p$-contour in either the upper or lower half-plane. \\

\noindent The poles in $p$ are located at
\begin{align}
    p&=\pm k_3 \nonumber \\
    p&=\pm(k_1 +k_2).
\end{align}
When evaluating the contour integral, we should recall that the Bunch-Davies boundary condition implies that the $k_a$ have a small, negative imaginary part. Finally, we can close the contour in the lower half-plane to find
\begin{equation}
    \mathcal{M}(\alpha,\beta) = {-\left({\rm Res}_{p=(k_1+k_2)}\bigg[{\cal N}(k_1,k_2,k_3,p)\bigg] + {\rm Res}_{p=k_3}\bigg[{\cal N}(k_1,k_2,k_3,s)\bigg]\right)}\,.
\end{equation}

\noindent The residue at $(k_1+k_2)$ can be computed for generic $n,m$ but the same is not true for the residue at $k_3$, for which we cannot find a closed-form expression. In any case we notice that, taken separately, the residues at $k_3$ and $(k_1+k_2)$ both present a divergence at $k_1+k_2=k_3$. It is only when combined that such divergence cancels, as it was expected from the analysis of Section~\ref{subsec:exchange_diagrams}. \\

\noindent Let us end this section by writing down some expressions for the time integrals of interest so we can see the expected properties explicitly. Since in all cases the result must be symmetric in the exchange of $k_{1}$ and $k_{2}$, we write the integrals in terms of the symmetric polynomials in two variables: $\hat{e}_{1}=k_{1}+k_{2}$, $\hat{e}_{2} = k_{1}k_{2}$. We still use $k_{T} = \hat{e}_{1}+k_{3}$. We have: 
\begin{align}
\mathcal{I}_{1,0} &= \frac{i k_{3}^3}{2 M_{\text{pl}}^2 k_{T}^5}(7\hat{e}_{1}^3 + 14 \hat{e}_{1}\hat{e}_{2}+11\hat{e}_{1}^2 k_{3}-2\hat{e}_{2}k_{3}+5\hat{e}_{1}k_{3}^2+k_{3}^3), \\
\mathcal{I}_{2,0} &= \frac{3 i H^2 k_{3}^3}{2M_{\text{pl}}^2 k_{T}^7}[15 \hat{e}_{1}^5 + 30 \hat{e}_{1}^3 \hat{e}_{2} +k_{3}(25 \hat{e}_{1}^4 - 30 \hat{e}_{1}^2 \hat{e}_{2}) \nonumber \\ &\;\;\;\;+k_{3}^2(26 \hat{e}_{1}^3 +34 \hat{e}_{1} \hat{e_{2}})+k_{3}^3(22 \hat{e}_{1}^2 -2 \hat{e}_{2})+7 k_{3}^4 \hat{e}_{1}+k_{3}^5], \\
\mathcal{I}_{0,1} &= \frac{i k_{3}}{2M_{\text{pl}}^2 k_{T}^5}(- \hat{e}_{1}^3-2\hat{e}_{1}\hat{e}_{2}+3 \hat{e}_{1}^2k_{3}+14 \hat{e}_{2}k_{3}+5\hat{e}_{1}k_{3}^2+k_{3}^3), \\
\mathcal{I}_{1,1}  &= \frac{i H k_{3}}{4 M_{\text{pl}}^2 k_{T}^5}(- \hat{e}_{1}^3-2\hat{e}_{1}\hat{e}_{2}+3 \hat{e}_{1}^2k_{3}+14 \hat{e}_{2}k_{3}+5\hat{e}_{1}k_{3}^2+k_{3}^3),  \\
\mathcal{I}_{2,1} &= \frac{12 i H^2 k_{3}^2}{M_{\text{pl}}^2 k_{T}^7}(\hat{e}_{1}^4+3\hat{e}_{1}^2 \hat{e}_{2} - 7 \hat{e}_{1}\hat{e}_{2}k_{3}-\hat{e}_{1}^2 k_{3}^2 + 2 \hat{e}_{2}k_{3}^2).
\end{align}
We note that the expressions for $\mathcal{I}_{0,1}$ and $\mathcal{I}_{1,1}$ are proportional which was to be expected. This is a consequence of the fact that for $m \neq 0$ and odd $n$, we can integrate by parts to write the quadratic mixing in terms of operators with lower values of $n$. As we explained before, it is still useful to keep our current labelling and definition of $n$ however, since we expect the contact contributions to be different for all $n$. For larger values of $n$, it will remain the case that for odd $n$ we can write the result in terms of expressions with lower and even $n$. As another example, we have $\mathcal{I}_{3,1} = \frac{9 H}{2}\mathcal{I}_{2,1} - 3 H^2 \mathcal{I}_{1,1}$. As expected, we see that only total-energy poles arise, and once we include the factors of $k_{3}^{2m}$ the MLT will be satisfied for each external leg.
\subsection{Contact diagram}

\noindent In addition to the exchange contributions we have just computed, for Type-I bispectra there are also contact diagrams that contribute to the bispectra at the same order in perturbation theory. These arise due to the cubic interactions in \eqref{2BBCT} which take the form
 \begin{align} \label{2BBCTInteractions}
S_{\gamma, \text{2BB}} \supset  \frac{M_{\text{pl}}^2}{4} \int d \eta d^3 x \sum_{n,m=1} g_{n,m} a^{q}(\eta) (\gamma_{ij})^{n+1}\left(- m\gamma^{lk} \partial^{2m-2} \partial_{l}\partial_{k} \gamma'_{ij} + \sum_{p=0}^{m-1} \partial^{2p} Q_{s}[\gamma, \partial^{2m-2 - 2p} \gamma']_{ij} \right),
 \end{align}
 where $q = 2-2m-n$ and $m \geq 1$.
 Each contribution contains one $\gamma_{ij}$ with $n+1$ time derivatives, one with a single time derivative, and one without any. Up to permutations and overall factors, the time integral that we therefore need to compute is
 \begin{align}
\mathcal{J}_{n,m}(k_{1},k_{2},k_{3}) = \int d \eta ~ \eta^{n+2m-2} K^{(n+1)}(k_{1}, \eta) K'(k_{2},\eta) K(k_{3},\eta).
 \end{align}
 With the result of this integral we can multiply it by the tensor structure dictated by the cubic interactions, then sum over permutations. Given that $n+2m-2 \geq 0$, the result of this integral will only have poles at $k_{T} = 0$, and no logs. Using \eqref{Kderivs} and
 \begin{align}
 \lim_{\eta \rightarrow 0} \int d \eta ~ \eta^{z} e^{i k_{T} \eta} = \frac{z! (-i)^{1-z}}{k_{T}^{1+z}} + \mathcal{O}(\eta^{z+1}),
 \end{align}
 for $z \geq 0$, we find that at late times the bulk time integral yields
 \begin{align}
\mathcal{J}_{n,m}^{123} &= (-1)^{m-n} i n (n+2m-1)! \frac{k_{1}^{n+1}k_{2}^2}{k_{T}^{n+2m}} \nonumber \\
& + (-1)^{1+m-n} i (n+2m)! \frac{k_{1}^{n+1}k_{2}^2(k_{1}- n k_{3})}{k_{T}^{n+2m+1}} \nonumber \\
& + (-1)^{1+m-n} i (n+2m+1)! \frac{k_{1}^{n+2}k_{2}^2 k_{3}}{k_{T}^{n+2m+2}},
 \end{align}
where we have introduced a more compact notation with $\mathcal{J}_{n,m}^{123} \equiv \mathcal{J}_{n,m}(k_{1},k_{2},k_{3})$. We see that the degree of the leading total-energy pole is equal to the number of derivatives in the cubic vertices, as expected \cite{BBBB}, and one can check that this expression satisfies the MLT for each external energy. \\

\noindent Now for the tensor structure we have 
 \begin{align}
Q_{s}[\gamma, \gamma']_{ij}  = - \partial_{i} \gamma_{kl} \partial_{k} \gamma'_{lj} + \partial_{l} \gamma_{jk} \partial_{k} \gamma'_{il} + (i \leftrightarrow j),
\end{align}
and in the final term in \eqref{2BBCTInteractions} we can integrate by parts to move all $\partial^{2p}$ terms onto the first $\gamma_{ij}$ such that 
 \begin{align} 
S_{\gamma, \text{2BB}} \supset  - & \frac{M_{\text{pl}}^2}{4} \int d \eta d^3 x \sum_{n,m=1} a^{q}(\eta) m g_{n,m} (\gamma_{ij})^{n+1} \gamma^{lk} \partial^{2m-2} \partial_{l}\partial_{k} \gamma'_{ij} \nonumber \\
+ &\frac{M_{\text{pl}}^2}{4} \int d \eta d^3 x \sum_{n,m=1} a^{q}(\eta) \sum_{p=0}^{m-1}  g_{n,m} \partial^{2p} (\gamma_{ij})^{n+1} Q_{s}[\gamma, \partial^{2m-2 - 2p} \gamma']_{ij}.
 \end{align}
From the first line, once we convert to momentum space using \eqref{gammaFT}, we will find tensor structures of the form $k_{2}^{2m-2} e_{ij}^{h_{1}}(\bfk_{1})e_{ij}^{h_{2}}(\bfk_{2}) k_{l}^{2}k_{k}^{2} e_{lk}^{h_{3}}(\bfk_{3})$, and if we convert this expression into one with spinors, for the $+++$ configuration we find
 \begin{align}
k_{2}^{2m-2} e_{ij}^{+}(\bfk_{1})e_{ij}^{+}(\bfk_{2}) k_{l}^{2}k_{k}^{2} e_{lk}^{+}(\bfk_{3})= \frac{1}{16} k_{2}^{2m-2} I_{3}^2 \text{SH}_{+++}.
 \end{align}
 We remind the reader that the $++-$ configuration can be extracted from this expression, as we explained above. For the second line, the tensor structures are of the form $k_{1}^{2p} k_{2}^{2m-2-2p}e_{ij}^{h_{1}}(\bfk_{1})[e_{il}^{h_{2}}(\bfk_{2})e_{jk}^{h_{3}}(\bfk_{3}) k_{k}^{2}k_{l}^{3} - e_{lj}^{h_{2}}(\bfk_{2})e_{kl}^{h_{3}}(\bfk_{3}) k_{k}^{2}k_{i}^{3}]$, which when converted to spinors for the $+++$ configuration yields
 \begin{align}
k_{1}^{2p} k_{2}^{2m-2-2p}e_{ij}^{+}(\bfk_{1})[e_{il}^{+}(\bfk_{2})e_{jk}^{+}(\bfk_{3}) k_{k}^{2}k_{l}^{3} - e_{lj}^{+}(\bfk_{2})e_{kl}^{+}(\bfk_{3}) k_{k}^{2}k_{i}^{3}] = -\frac {1}{8}k_{1}^{2p} k_{2}^{2m-2-2p} k_{3} I_{3}  \text{SH}_{+++}.  
 \end{align}
 We can now collect everything together. By multiplying these tensor structures by the result of the time integral, and including all constant factors as dictated by the Feynman rules we reviewed in Section \ref{Generalities}, for the $+++$ helicity configuration we find
 \begin{align}
&\psi_{\text{2BB,contact}}^{+++} = \frac{i M^2_{\rm pl}}{64} \text{SH}_{+++} \sum_{n,m=1} (-1)^{n+m} g_{n,m}  H^{n+2m-2}  m \left[ k_{2}^{2m-2}I_{3}^2 \mathcal{J}_{n,m}^{123} + 5 ~ \text{perms} \right] \nonumber \\
& +  \frac{i M^2_{\rm pl} }{16} \text{SH}_{+++} \sum_{n,m=1} (-1)^{n+m} g_{n,m} H^{n+2m-2}  \sum_{p=0}^{m-1} \left[ k_{1}^{2p} k_{2}^{2m-2p-2}k_{3} I_{3} \mathcal{J}_{n,m}^{123} + 5 ~ \text{perms} \right].
 \end{align}
 We have summed over the remaining permutations to find an object with the correct symmetry. The symmetrisation in the expression for $Q_{s}$ simply yields a factor of $2$. Now to extract the wavefunction coefficient for the $++-$ helicity configuration we need to send $k_{3} \rightarrow -k_{3}$ and $\text{SH}_{+++} \rightarrow \text{SH}_{++-}$ for each term, while keeping $\mathcal{J}_{n,m}$ fixed. This yields
  \begin{align}
&\psi_{\text{2BB,contact}}^{++-} = \frac{i M^2_{\rm pl} }{64} \text{SH}_{++-} \sum_{n,m=1} (-1)^{n+m} g_{n,m}  H^{n+2m-2} m [k_{2}^{2m-2}k_{T}^2 \mathcal{J}_{n,m}^{123} + k_{1}^{2m-2}k_{T}^2 \mathcal{J}_{n,m}^{213} \nonumber \\
&+k_{1}^{2m-2}I_{1}^2 \mathcal{J}_{n,m}^{312} + k_{3}^{2m-2}I_{1}^2 \mathcal{J}_{n,m}^{132}+k_{2}^{2m-2}I_{2}^2 \mathcal{J}_{n,m}^{321} +  k_{3}^{2m-2}I_{2}^2 \mathcal{J}_{n,m}^{231}] \nonumber \\
& +  \frac{i M^2_{\rm pl} }{16} \text{SH}_{++-} \sum_{n,m=1} (-1)^{n+m}g_{n,m}  H^{n+2m-2} \sum_{p=0}^{m-1}[-k_{1}^{2p} k_{2}^{2m-2p-2}k_{3}k_{T} \mathcal{J}_{n,m}^{123} \nonumber \\
&-k_{2}^{2p} k_{1}^{2m-2p-2}k_{3}k_{T} \mathcal{J}_{n,m}^{213} 
 - k_{1}^{2p}k_{3}^{2m-2p-2}k_{2}I_{1} \mathcal{J}^{132}_{n,m} \nonumber \\ & - k_{3}^{2p}k_{1}^{2m-2p-2}k_{2}I_{1} \mathcal{J}^{312}_{n,m} - k_{2}^{2p}k_{3}^{2m-2p-2}k_{1}I_{2} \mathcal{J}_{n,m}^{231} - k_{3}^{2p}k_{2}^{2m-2p-2}k_{1}I_{2}\mathcal{J}_{n,m}^{321}].
 \end{align}
 We note that for both helicity configurations these contributions to the wavefunction are real since the overall factor of $i$ is cancelled by the $i$ in $\mathcal{J}_{n,m}$. This ensures that these contributions do indeed contribute to the bispectra since for parity-even interactions only the real part contributes (c.f. \eqref{BispectrumEqn}).

\subsection{Putting everything together}

\noindent Now that we have all contributions to the cubic wavefunction coefficient, to leading order in new couplings, we can now convert these into expressions for the bispectra. In perturbation theory expectation values are algebraically related to wavefunction coefficients with the relations derived in a number of places e.g. \cite{Cabass:2021fnw,WFCtoCorrelators2,WFCtoCorrelators1}. When there is a small correction to the two-point function, as is the case here, the general expression is given by \eqref{WFtoCorrelator} which for parity-even interactions becomes 
\begin{align}
B_{3}^{\{ \lambda_{i} \}} = \frac{1}{\Pi_{i=1}^3 2 \text{Re} \left(\psi_{2,\text{GR}}\right)} \left[- 2 \text{Re} \left(\psi_{3,\text{total}}^{\{ \lambda_{i} \}}\right) +2 \text{Re} \left(\psi_{3,\text{GR}}^{\{ \lambda_{i} \}}\right)\left(\frac{\text{Re} \left(\delta \psi_{2}^{\lambda_{1}} \right)}{\text{Re}\left( \psi_{2,\text{GR}}\right)} + \text{2 perms} \right) \right],
\end{align}
where $\psi_{2,\text{GR}}$ is the GR contribution to the two-point function, while $\delta \psi_{2}$ is a small correction due to our higher-derivative corrections to the quadratic action. Since we are working up to linear order in new couplings, we take the first term in the square brackets to be all contributions we have computed in this section i.e. $\psi_{3,\text{total}} = \psi_{\text{2BB,exchange}}+\psi_{\text{2BB,contact}}$, while the $\psi_{3}$ in the second term must be the GR contribution since $\delta \psi_{2}$ is already linear the new couplings. In addition to $\psi_{3,\text{total}}$, we now also need the GR wavefunction up to cubic order, and the small corrections to $\psi_{2}$. \\

\noindent We have essentially already computed the GR cubic wavefunction coefficient when we computed the Type-I exchange diagrams so let us simply write the result here. We have\footnote{As we did with the quadratic wavefunction, here we are dropping any imaginary contributions to the GR cubic wavefunction coefficient.}
\begin{align}
\psi^{+++}_{3,\text{GR}} = \frac{M_{\text{pl}}^2}{64 H^{2}}\text{SH}_{+++} \frac{k_{T}^2}{k_{T}^2}(e_{3}+k_{T}e_{2}-k_{T}^3), \\
\psi^{++-}_{3,\text{GR}} = \frac{M_{\text{pl}}^2}{64 H^{2}}\text{SH}_{++-} \frac{I_{3}^2}{k_{T}^2}(e_{3}+k_{T}e_{2}-k_{T}^3). 
\end{align}
For the corrections to the two-point function, we need to compute a Feynman diagram that is analogous to the one in Figure \ref{figure:Contact} but with two rather than three external legs. Since this is a small correction to the quadratic wavefunction, we compute it in the way we compute any contact diagram: we insert a bulk-boundary propagator for each external line, add tensor structures and time derivatives as dictated by corresponding the bulk vertex, and use the Feynman rules we discussed above. For example, for $m = 0$ we have
\begin{align}
\delta \psi_{2}^{h h'} &= - i \sum_{n=1} \frac{g_{n,0} M_{\text{pl}}^2 H^{2n-2}}{4}   \int d \eta ~ \eta^{2n-2} K^{(2n+1)}_{\gamma}(k, \eta) K'_{\gamma}(k, \eta) \times 4 \delta_{h h'} \times 2 \\
 &= \sum_{n=1} \frac{(2n)!}{4^n} g_{n,0} M_{\text{pl}}^2 H^{2n-2} k^3 \delta_{h h'},
\end{align}
where we have used momentum conservation and summed over the two possible permutations. The computation for $m \neq 0$ is very similar and in total we have 
\begin{align}
\delta \psi_{2}^{h h'} = &\sum_{n=1} \frac{(2n)!}{4^n} g_{n,0} M_{\text{pl}}^2 H^{2n-2} k^3 \delta_{h h'} \nonumber \\ &+ \sum_{n=0,m=1} \frac{(2m+n-1)! (n-2m)}{2^{n+2m}} g_{n,m} M_{\text{pl}}^2 H^{n+2m-2} k^3  \delta_{h h'}.
\end{align}
We now have all the ingredients to compute the bispectra. Below we write down their final forms at linear order in the new couplings for a few choices of $n$ and $m$. We define
\begin{align}
\delta P = - \frac{ P_{\text{GR}} \delta \psi_{2}}{\psi_{2,\text{GR}}} = -\frac{H^4}{2M_{\text{pl}}^4 k^6} \delta \psi_{2}.
\end{align}
We have, for the examples with the lowest degree leading total-energy poles, 
\begin{align}
\text{GR}: & \quad  B_3^{+++} = \frac{H^4}{256 M_{\text{pl}}^4 e_{3}^3} \text{SH}_{+++}(k_{T}^3-k_{T}e_{2}-e_{3}) \label{GRB3} \\
                    & \quad   B_3^{++-} = \frac{H^4}{256 M_{\text{pl}}^4 e_{3}^3} \text{SH}_{++-} \frac{I_{3}^2}{k_{T}^2}(k_{T}^3-k_{T}e_{2}-e_{3}), \\
                    & \quad P_{\text{GR}}^{h h'}  =  \frac{H^2}{2 M_{\text{pl}}^2 k^3} \delta_{h h'}. \label{GRPS} \\
    n = 1, m = 0: & \quad  \delta B_3^{+++} = -\frac{H^6 g_{1,0} }{256 M_{\text{pl}}^4 e_3^3} \text{SH}_{+++} \frac{k_T^6 - k_T^4 e_2 - k_T^3 e_3  + 24 e_3^2}{k_T^3}, \\
                    & \quad  \delta B_3^{++-} = -\frac{H^6 g_{1,0} }{256 M_{\text{pl}}^4 e_3^3} \text{SH}_{++-} \frac{I_{3}^2}{k_{T}^2} \frac{k_T^6 - k_T^4 e_2 - k_T^3 e_3  + 24 e_3^2}{k_T^3}, \\
                    & \quad \delta P^{h h'}  = -  \frac{H^4 g_{1,0}}{4 M_{\text{pl}}^2 k^3} \delta_{h h'}. \\
                     n = 2, m = 0:  & \quad  \delta B_3^{+++} = - \frac{3 H^8 g_{2,0}}{256 M_{\text{pl}}^4 e_3^3} \text{SH}_{+++} \frac{k_T^8  - e_2 k_T^6  - e_3 k_T^5  +80 e_3^2 k_T^2 - 240 e_2 e_3^2}{k_{T}^5}
                    , \\
                    & \quad  \delta B_3^{++-} = - \frac{3 H^8 g_{2,0}}{256 M_{\text{pl}}^4 e_3^3} \text{SH}_{++-}\frac{I_{3}^2}{k_{T}^2} \frac{k_T^8  - e_2 k_T^6  - e_3 k_T^5  +80 e_3^2 k_T^2 - 240 e_2 e_3^2}{k_{T}^5}, \\
                    & \quad \delta P^{h h'} =  -  \frac{3 H^6 g_{2,0}}{4 M_{\text{pl}}^2 k^3} \delta_{h h'}. \\
                     n = 0, m = 1:  & \quad  \delta B_3^{+++} =   \frac{H^6 g_{0,1}}{256 M_{\text{pl}}^4 e_3^3} \text{SH}_{+++} \frac{k_T^6 -  k_T^4 e_2  -  k_T^3 e_3  + 72 e_3^2}{k_T^3} \\ 
                    & \quad  \delta B_3^{++-} = \frac{H^6 g_{0,1}}{256 M_{\text{pl}}^4 e_3^3} \text{SH}_{++-} \frac{\text{Poly}_{8a}(\hat{e}_{1},\hat{e}_{2},k_{3})}{k_{T}^5} \\
                    & \quad \delta P^{h h'} =   \frac{H^4 g_{0,1}}{4 M_{\text{pl}}^2 k^3} \delta_{h h'}.
                    \\
                     n = 1, m = 1:  & \quad  \delta B_3^{+++} =  \frac{H^7 g_{1,1}}{512 M_{\text{pl}}^4 e_3^3} \text{SH}_{+++} \frac{k_T^8 -  k_T^6 e_2  -  k_T^5 e_3  + 24 k_{T}^2 e_3^2 + 384 e_{2}e_{3}^2}{k_T^5} \\ 
                    & \quad  \delta B_3^{++-} = \frac{H^7 g_{1,1}}{512 M_{\text{pl}}^4 e_3^3} \text{SH}_{++-} \frac{\text{Poly}_{8b}(\hat{e}_{1},\hat{e}_{2},k_{3})}{k_{T}^5}  \\
                    & \quad \delta P^{h h'} =  \frac{H^5 g_{1,1}}{8 M_{\text{pl}}^2 k^3} \delta_{h h'},
\end{align}
where we have defined the polynomials
\begin{align}
\text{Poly}_{8a}(\hat{e}_{1},\hat{e}_{2},k_{3}) &= \hat{e}_{1}^8 + 3 \hat{e}_{1}^7 k_{3}-\hat{e}_{1}^6(\hat{e}_{2}-2 k_{3}^2)-3\hat{e}_{1}^5 k_{3}(\hat{e}_{2}+ k_{3}^2) - 6 \hat{e}_{1}^4 k_{3}^4 + \hat{e}_{1}^3(6 \hat{e}_{2} k_{3}^3-3 k_{3}^5) \nonumber \\& +\hat{e}_{1}^2(72\hat{e}_{2}^2 k_{3}^2+3 \hat{e}_{2} k_{3}^4 + 2 k_{3}^6)+3\hat{e}_{1}(80\hat{e}_{2}^2 k_{3}^3-\hat{e}_{2} k_{3}^5 + k_{3}^7)+(72 \hat{e}_{2}^2 k_{3}^4 - 2 \hat{e}_{2}k_{3}^6 + k_{3}^8), \\
\text{Poly}_{8b}(\hat{e}_{1},\hat{e}_{2},k_{3}) &= \text{Poly}_{8a}(\hat{e}_{1},\hat{e}_{2},k_{3}) - 48\hat{e}_{2}^2 k_{3}^2(\hat{e}_{1}^2 - 8 \hat{e}_{2}+6 \hat{e}_{1}k_{3}+k_{3}^2),
\end{align}
using the symmetric polynomials in two variables: $\hat{e}_{1} = k_{1}+k_{2}$, $\hat{e}_{2} = k_{1}k_{2}$. \\

\noindent Perhaps surprisingly we see that for these four examples the bispectra for both helicity configurations can be written as a linear sum of the GR bispectra and those that we will introduce in the following section, namely, the Type-II bispectra. Indeed, once we subtract the GR contributions from these Type-I bispectra, we are left with contributions that contain an overall $e_{3}^2$ which is the defining property of Type-II bispectra as we will show. This does not mean that the two building blocks operators are completely degenerate with the three building block ones since observables could still differ at higher order in perturbations e.g. the corresponding trispectra could be different. However, this observation certainly motivates us to construct these bispectra directly using consistency relations which may shed light on why this apparent degeneracy arises. Our result here suggests that if there is no correction to the power spectrum, then to satisfy the consistency relations we require an overall $e_{3}^2$ and it would be nice to prove this in generality. We will come back to this in the future.

\subsection{Checking the consistency relations}

Given that we are working within the EFToI, our bispectra should satisfy the consistency relations of single-clock cosmologies which relate the soft limit of $n$-point functions to lower point ones \cite{Hinterbichler:2013dpa,Maldacena:2002vr,Creminelli:2012ed}. They arise from the unbroken spatial diffeomorphisms and offer a good consistency check of our results. The leading order graviton soft theorem is
\be
\langle \gamma^{h_1}_{\vb{q}} \gamma^{h_2}_{\bfk -\vb{q}/2} \gamma^{h_3}_{-\bfk -\vb{q}/2} \rangle' \sim \frac{3}{2} (\epsilon_{ij}^{h_1}(\vb{q}) \hat{\bfk}_i \hat{\bfk}_j ) P^{h_1 h_1}_{\gamma}(q) P^{h_2 h_3}_{\gamma}(k) \quad \text{as} \quad \frac{q}{k} \to 0,
\ee
where we have introduced the notation $\langle \ldots \rangle'$ to denote a correlator with the momentum-conserving delta function stripped off. \\

\noindent We will focus on the consistency relation for the $+++$ configuration which is enough to verify that our results are correct. We first need to compute the tensor structures that appear on both sides of the soft theorem. In both cases we can write these solely in terms of the three external energies by picking a basis for the vectors and polarisation tensors. Using momentum conservation and $SO(3)$ invariance, we can make each of the three external vectors lie in the $(x,y)$ plane and we choose $\bfk_{1} = k_{1}(1,0,0)$ without loss of generality. We can then write the corresponding polarisation tensor as
\begin{align}
e^{\pm}(\bfk_{1}) & =  \begin{pmatrix}
0 & 0 & 0 \\
0 & 1 & \pm i \\
0 & \pm i & -1 
\end{pmatrix} \,,
\end{align}
which is traceless, transverse to the momentum and has the correct normalisation. The other vectors and polarisation tensors can easily be extracted from these by performing rotations and using momentum conservation (see e.g. \cite{Cabass:2021fnw}). It is then straightforward to see that\footnote{Despite the overall factor of $e_{3}^{-2}$, this object is not singular in the soft limit thanks to the tunings between terms in the numerator.} 
\begin{align}
\text{SH}_{+++} = - 8 e^{h_{1}}_{ij}e^{h_{2}}_{jk} e^{h_{3}}_{ki} =  -  \frac{k_T^3}{ e_3^2} \left( 8 e_3 - 4 k_T e_2+ k_T^3 \right)\,,
\end{align}
with a similar expression for $\text{SH}_{++-}$ \cite{Cabass:2021fnw}.\footnote{We note that the polarisation tensors are matrices of rank $1$ which ensures that we can use the spinor helicity formalism without loss of generality. We can therefore use this expression for $\text{SH}_{+++}$ to easily write all tensor structures in terms of the energies.} Similarly, we have
\begin{align}
e_{ij}^{h_{1}}\hat{k}_{i}^2 \hat{k}_{j}^3 = -\frac{k_{2}}{k_{3}} \left(1 - \frac{(k_{1}^2+k_{2}^2-k_{3}^2)^2}{4 k_{1}^2 k_{2}^2} \right).
\end{align}
With these expressions, and the bispectrum and power spectrum of GR given in \eqref{GRB3} and \eqref{GRPS} respectively, we can easily see that the soft theorem is satisfied in pure gravity. Using the general consistency relation at zeroth order in the new EFToI couplings, we can then obtain a simplified relation at first order in terms of the new bispectra and change in the power spectrum given by
\be
\delta B^{+++}(q, |\bfk - \vb{q}/2|, |\bfk + \vb{q}/2|)  \sim 2 B_{\text{GR}}^{+++}(q, |\bfk - \vb{q}/2|, |\bfk + \vb{q}/2|)  \frac{\delta P^{++}_{\gamma}(k)}{P^{++}_{\gamma, \text{GR}}(k)}. 
\ee
It is then simple to see that this relationship holds for the examples we wrote above at leading order in the soft momentum i.e. at $\mathcal{O}(1 / q^3)$. We have also checked that the $++-$ configuration satisfies the appropriate soft limits. If we take $\vec{k}_{3}$ soft then the bispectrum contributes to the left-hand side of the soft theorem and the correction to the power spectrum ensures that it is satisfied, while if we take either $\vec{k}_{1}$ or $\vec{k}_{2}$ soft then to leading order there is no contribution to the left-hand side of the correlator which is comforting since the right-hand side would only be non-zero at leading order if the $+$ and $-$ modes were correlated, which is not the case. In checking this we see the welcome appearance of $I_{3}^2$ in the $++-$ bispectrum: at leading order in the soft momentum ($\vec{k}_{1}$ or $\vec{k}_{2}$) this combination of energies vanishes thereby ensuring that there is no contribution to the left-hand side of the soft theorem. 



\section{Type-II bispectra} \label{TypeII}

\noindent Let's now turn our attention to Type-II bispectra that come from three building block operators. In this case our Lagrangian contains all tensor structures derived in \cite{Cabass:2021fnw} with the freedom to add additional derivatives. The only constraint coming from the fact these are EFToI operators is that each $\gamma_{ij}$ must come with at least one time derivative, as we showed in Section \ref{GravitonInteractions}. This makes sense from the point of view of symmetries: the theory should be invariant under spatial diffeomorphisms and if we write these symmetry transformations as a Taylor expansion in $x^{i}$, $\gamma_{ij}$ (and $\zeta$) \cite{Hinterbichler:2013dpa}, then the operators with the fewest powers of $\gamma_{ij}$ must be invariant under the field-independent part of the symmetry transformation which would simply be a sum of polynomials in $x^{i}$ under which $\gamma'_{ij}$ is invariant. Here there are no corrections to the power spectrum, and therefore the cubic terms are those with the fewest powers of $\gamma_{ij}$. \\

\noindent Now as was explained in detail in \cite{Cabass:2021fnw}, to construct bispectra we take one of the allowed tensor structures and multiply it by a solution to the MLT before summing over permutations. In terms of polarisation tensors and spatial momenta, the five allowed tensor structures, up to permutations, are (these also follow from the action in \eqref{3BBCT})
\begin{align} \label{TensorStructures}
\alpha=0: ~~ & e_{ij}^{h_{1}}e^{h_{2}}_{jk}e^{h_{3}}_{ki}, \\
\alpha=2: ~~ & e_{lm}^{h_{1}}e_{lm}^{h_{2}}e_{ij}^{h_{3}}k_{1}^i k_{2}^j ~~~ \text{and} ~~~ e_{lm}^{h_{1}}e_{il}^{h_{2}}e_{jm}^{h_{3}}k_{1}^i k_{1}^j, \\
\alpha=4: ~~ & e_{lk}^{h_{1}}e^{h_{2}}_{mk}e^{h_{3}}_{ij} k_{1}^i k_{2}^j k_{3}^l k_{3}^m, \\
\alpha=6: ~~ & e_{il}^{h_{1}}e^{h_{2}}_{jm}e^{h_{3}}_{kn} k_{1}^m k_{1}^k k_{2}^i k_{2}^n k_{3}^l k_{3}^j. 
\end{align}
As we discussed in the previous section, wavefunction coefficients and bispectra are actually more compactly presented by converting to the spinor helicity formalism. In this case a general parity-even wavefunction coefficient for the $+++$ configuration can be written as
\begin{align}
\psi_{3}^{+++}(\{k \}, \{ \bfk \}) = \text{SH}_{+++} \sum_{\text{permutations}} h_{\alpha}(\{ k \}) \psi_{3}^{\text{trimmed}}(\{ k \}),
\end{align}
where the parity-even choices for $h_{\alpha}$ are \cite{Cabass:2021fnw}
\begin{align}
h_{0} &= 1, \\
h_{2} &= k_{1}^2 ~~~ \text{and} ~~~ k_{1}k_{2}, \\
h_{4} &= I_{1}^2 I_{2}I_{3}, \\
h_{6} &= I_{1}^2 I_{2}^2 I_{3}^2,
\end{align}
and we remind the reader that the symmetries of the trimmed part of the wavefunction are dictated by $h_{\alpha}$, and $I_{a} = k_{b}+k_{c} - k_{a}$, with $a\neq b \neq c$. The trimmed wavefunction is the contribution that comes from time evolution. These five structures follow directly from converting the tensor structures in \eqref{TensorStructures} into spinor variables, however there is a further simplification that was noted in \cite{Cabass:2021fnw}. Consider the $k_{1}^2$ possibility for $h_{2}$. In the bispectra this factor would be multiplied by a solution to the MLT and then summed over permutations. However, $k_{1}^2$ is by itself also a solution to the MLT so in fact this $\alpha=2$ possibility is already captured by $\alpha=0$. Here we are aiming to construct a complete set of Type-II bispectra so we can therefore work with the restricted set
\begin{align}
h_{0} &= 1, \\
h_{2} &= k_{1}k_{2} , \\
h_{4} &= I_{1}^2 I_{2}I_{3}, \\
h_{6} &= I_{1}^2 I_{2}^2 I_{3}^2,
\end{align}
with only a single possibility for each $\alpha$. From these expressions we can extract the $++-$ configuration, as we explained in Section \ref{SHF}. We have
\begin{equation}
\label{eq:psi++-}
\psi^{++-}_3(\{k \}, \{ \bfk \}) = \text{SH}_{++-} \sum_{\rm permutations} h_{\alpha}(k_1, k_2, -k_3) \psi^{\rm trimmed}_3(\{ k \}) \,.
\end{equation}

\noindent Now the trimmed wavefunction can be a rational function of the three energies, can contain a log divergence in $- \eta_{0}k_{T}$ and can also contain poles at $\eta_{0} = 0$ of at most cubic degree. These allowed structures follow from the combination of scale invariance and the assumption of a Bunch-Davies vacuum \cite{Cabass:2021fnw,BBBB}. As we mentioned above, in our EFToI operators there are too many derivatives for log divergences to arise, and any $\eta_{0} = 0$ poles drop out of the correlator, which can be shown in complete generality using a combination of the MLT and the COT \cite{Cabass:2021fnw}. They drop out since their coefficients are always imaginary and only real wavefunction coefficients contribute to the bispectra. In any case, the fact that each $\gamma_{ij}$ is differentiated at least once is enough to rule out $\eta_{0} = 0$ poles in the wavefunction. For our interests we can therefore restrict the trimmed wavefunction to be a rational function with poles only occurring when $k_{T} = 0$. We therefore have
\begin{align}
\psi_{3}^{\text{trimmed}}(\{ k \}) = \frac{1}{k_{T}^p} \text{Poly}_{3+p-\alpha}(k_{1},k_{2},k_{3}),
\end{align}
with $p$ the order of the leading total-energy pole, and $3+p-\alpha$ the degree of the polynomial which is fixed by scale invariance. \\

\noindent Now, given the form of \eqref{3BBCT}, in our case the trimmed wavefunction would arise from bulk time integrals of the form
\begin{align}
\int d \eta ~ \eta^{-r} K^{(1+n_{1})}(k_{1},\eta) K^{(1+n_{2})}(k_{2},\eta) K^{(1+n_{3})}(k_{3},\eta), 
\end{align}
with each bulk-boundary propagator differentiated at least once and $r \leq 1$. We have 
\begin{align}
K'(k, \eta) = k^2 \eta e^{i k \eta},
\end{align}
so each external energy will appear at least quadratically in $\psi_{3}^{\text{trimmed}}$. We can therefore update our ansatz to\footnote{This overall factor of $e_{3}^2$ is enough to rule out $\eta_{0} = 0$ poles since by scale invariance any negative powers of $\eta_{0}$ would also require negative powers of $k_{T}$ but no such structures are allowed \cite{BBBB}.} 
\begin{align} \label{TrimmedFinal}
\psi_{3}^{\text{trimmed}}(\{ k \}) = \frac{e_{3}^2}{k_{T}^p} \text{Poly}_{p-3-\alpha}(k_{1},k_{2},k_{3}),
\end{align}
which by virtue of the factor of $e_{3}^2$ satisfies the MLT for each external energy and for all choices of $\text{Poly}_{p-3-\alpha}(k_{1},k_{2},k_{3})$. In \cite{Cabass:2021fnw} it was shown that all solutions to the MLT come from a bulk time integral, so \eqref{TrimmedFinal} reproduces precisely the desired trimmed wavefunctions that follow from \eqref{3BBCT}. This can also been seen from the fact that $K'(k,\eta) = k^2 K_{\text{CC}}(k, \eta)$, where $K_{\text{CC}}$ is the bulk-boundary propagator for a conformally coupled scalar. The MLT for such a field is trivial \cite{MLT}, so all rational functions are admissible and can be generated by taking the necessary time derivatives of $K_{\text{CC}}$. Without having to compute any time integrals we can now construct $\psi_3^{\text{trimmed}}$ for each $\alpha$, and use these expressions to compute the bispectra $B_{\text{3BB}, \alpha}^{+++}$ and $B_{\text{3BB}, \alpha}^{++-}$. Note that when we convert wavefunction coefficients to correlators we pick up a factor of $1/e_{3}^3$ from the inverse powers of $\psi_{2}$ in \eqref{BispectrumEqn}, and in the following we absorb all constant factors, such as $H$ and $M_{\text{pl}}$, into the arbitrary polynomials.   \\

\noindent Let's start with $\alpha=0$ where the trimmed wavefunction should be fully symmetric in the external energies since the tensor structure is. We can therefore write the polynomial as a function of $k_{T}$, $e_{2}$ and $e_{3}$. The general bispectra are then
\begin{align}
 B_{\text{3BB}, 0}^{+++} &= \frac{e_{3}^2 \text{SH}_{+++}}{e_{3}^3 k_{T}^p} \text{Poly}_{p-3}(k_{T},e_{2},e_{3}), \\
 B_{\text{3BB}, 0}^{++-} &= \frac{e_{3}^2 \text{SH}_{++-}}{e_{3}^3 k_{T}^p}  \text{Poly}_{p-3}(k_{T},e_{2},e_{3}),
\end{align}
where in all cases the degree of the polynomial must be a non-negative number so for $\alpha=0$ we need $p \geq 3$. Now for $\alpha=2$ we have $h_{2} = k_{1}k_{2}$, so the trimmed wavefunction only needs to be symmetric in the exchange of $k_{1}$ and $k_{2}$. We therefore have 
\begin{align}
B_{\text{3BB}, 2}^{+++} &=  \frac{e_{3}^2 \text{SH}_{+++}}{e_{3}^3 k_{T}^p}[k_{1}k_{2}\text{Poly}_{p-5}(k_{T},e_{2},k_{3})+k_{1}k_{3}\text{Poly}_{p-5}(k_{T},e_{2},k_{2})+k_{2}k_{3}\text{Poly}_{p-5}(k_{T},e_{2},k_{1})], \\
B_{\text{3BB}, 2}^{++-} &=  \frac{e_{3}^2 \text{SH}_{++-}}{e_{3}^3 k_{T}^p}[k_{1}k_{2}\text{Poly}_{p-5}(k_{T},e_{2},k_{3})-k_{1}k_{3}\text{Poly}_{p-5}(k_{T},e_{2},k_{2})-k_{2}k_{3}\text{Poly}_{p-5}(k_{T},e_{2},k_{1})].
\end{align}
We would naturally write the arguments of the polynomial, for the $k_{1}k_{2}$ permutation, as $\hat{e}_{1} = k_{1}+k_{2}$, $\hat{e}_{2} = k_{1}k_{2}$ and $k_{3}$ given its symmetries but we can replace $\hat{e}_{1}$ with $k_{T}$ and $\hat{e}_{2}$ with $e_{2}$ without loss of generality since $k_{T} = \hat{e}_{1}+k_{3}$ and $e_{2} = \hat{e}_{2} + k_{3}\hat{e}_{1}$. The case of $\alpha=4$ is very similar to $\alpha=2$ since the symmetries of $h_{4}$ are the same as $h_{2}$. We have  
\begin{align}
B_{\text{3BB}, 4}^{+++} &=  \frac{e_{3}^2 \text{SH}_{+++}}{e_{3}^3 k_{T}^p}[I_{3}^2I_{1}I_{2}\text{Poly}_{p-7}(k_{T},e_{2},k_{3})+I_{2}^2I_{1}I_{3}\text{Poly}_{p-7}(k_{T},e_{2},k_{2})+I_{1}^2I_{2}I_{3}\text{Poly}_{p-7}(k_{T},e_{2},k_{1})], \\
 B^{++-}_{\text{3BB},4} &= \frac{e_{3}^2\text{SH}_{++-}}{e_{3}^3k_{T}^p} [k_{T}^2 I_{1}I_{2}\text{Poly}_{p-7}(k_{T}, e_{2}, k_{3}) - I_{1}^2 I_{2}k_{T}\text{Poly}_{p-7}(k_{T}, e_{2}, k_{2}) - I_{2}^2 I_{1}k_{T}\text{Poly}_{p-7}(k_{T}, e_{2}, k_{1})],
\end{align}
where we have used that under $k_{3} \rightarrow -k_{3}$ we have $I_{3} \rightarrow k_{T}$ and $I_{1} \rightarrow - I_{2}$. Finally, for $\alpha=6$ we again have a symmetric $h_{6}$ so the bispectra are 
\begin{align}
 B_{\text{3BB}, 6}^{+++} &= \frac{e_{3}^2 \text{SH}_{+++}}{e_{3}^3 k_{T}^p} I_{1}^2I_{2}^2I_{3}^2\text{Poly}_{p-9}(k_{T},e_{2},e_{3}), \\
 B_{\text{3BB}, 6}^{++-} &= \frac{e_{3}^2 \text{SH}_{++-}}{e_{3}^3 k_{T}^p}  I_{1}^2I_{2}^2 k_{T}^2\text{Poly}_{p-9}(k_{T},e_{2},e_{3}).
\end{align}
The above structures give the most general graviton bispectra coming from three building block operators in the EFToI, to all orders in derivatives (or equivalently to all orders in $p$). Compared to the general Lagrangian \eqref{3BBCT}, the resulting bispectra take a very compact form. \\

\noindent For Type-I bispectra we checked that the consistency relations of the EFToI are satisfied by our results. This provides a non-trivial check. For these Type-II bispectra we see that the consistency relation is clearly satisfied: there is no correction to the power spectrum since these shapes come from three building block operators and indeed these shapes do not contribute in the leading soft limit thanks to the overall factor of $e_{3}^2$. So the consistency relation is also satisfied here. These Type-II bispectra therefore represent the most general tree-level subset that satisfy the consistency relation without a need to correct the graviton power spectrum. 

\section{Conclusions and outlook} \label{Conclusions}

In this paper we have computed late-time inflationary three-point functions for massless gravitons in the Effective Field Theory of Inflation (EFToI). At tree-level, and to leading order in the field theory couplings, there are two Feynman diagrams that contribute to the cubic wavefunction coefficient of massless gravitons: one is a contact diagram due to cubic self-interactions of the graviton, while the second is an exchange diagram that perturbatively accounts for possible corrections to the graviton power spectrum. Computationally we have concentrated on these wavefunction coefficients, but have used standard techniques to extract expectation values, namely bispectra, from these objects.   \\

\noindent We have, for the first time, shown that the quadratic and cubic action for massless gravitons, that appears in addition to the Einstein-Hilbert part, can be derived by considering covariant operators constructed out of the extrinsic curvature only. We arrived at this conclusion by performing various field redefinitions to eliminate operators that contain the three-dimensional Ricci tensor. At the level of covariant operators we have distinguished between those that are quadratic or cubic in the extrinsic curvature. The former contribute to both the quadratic and cubic operators for the transverse traceless fluctuation, and we refer to the corresponding bispectra as Type-I, while the latter only contribute to the cubic operators and we refer to these bispectra as Type-II. In both cases we computed these bispectra to all orders in derivatives, and have shown that our results are a consistent sub-set of the general graviton bispectra constructed in \cite{Cabass:2021fnw}. \\

\noindent For Type-I bispectra, both types of Feynman diagrams contribute and they are tied together by spatial diffeomorphisms and the non-linear realisation of time diffeomorphisms. For the exchange diagram the leading order contribution comes from taking the cubic interaction to be that of GR and with the quadratic mixing coming from expanding two building block operators to quadratic order. We use the techniques of \cite{Meltzer:2021zin} to efficiently compute the necessary bulk time integrals. The contact diagram arises from the self-interactions coming from expanding two building block operators to cubic order. Such interactions only arise in the presence of spatial derivatives: in the absence of spatial derivatives the symmetries of the extrinsic curvature ensure that there are no cubic corrections. In this case we explicitly compute the necessary bulk time integrals. Both diagrams contribute to the bispectra and we have checked that our bispectra satisfy the leading order consistency relation of the EFToI.  \\

\noindent For Type-II bispectra, only contact diagrams contribute since there are no corrections to the quadratic action. In this case we have used the techniques of \cite{Cabass:2021fnw} to write down the most general allowed wavefunction coefficients. Since each contribution to the cubic action contains three gravitons differentiated with respect to conformal time at least once, the wavefunction coefficients always contain an overall factor of $(k_{1}k_{2}k_{3})^2$ which ensures that the MLT \cite{MLT} is trivially satisfied. We have again extracted bispectra from our wavefunction expressions and the overall factor of $(k_{1}k_{2}k_{3})^2$ ensures that the leading order consistency relations are again satisfied. \\

\noindent The bispectra that we have derived should be taken in addition to those of GR which were first computed in \cite{Maldacena:2011nz}. All of the corrections we have derived have leading order total-energy poles that are of a higher degree than that of pure gravity which makes sense since they all come from operators with more than two derivatives. A consequence of this is that pure gravity is the only case where the bispectrum for $+++$ configuration does not have a total-energy pole. The Type-I bispectra will always have a smaller amplitude compared to their GR counter-parts since we treated the corrections to the power spectra perturbatively, however the restrictions on the size of Type-II bispectra are weaker since there is no correction to the two-point function. \\

\noindent There are avenues for future work:

\begin{itemize}
    \item In this work we have been using a combination of bulk and bootstrap tools. It would be great to be able to derive this collection of EFToI graviton bispectra purely using bootstrap methods. Consistency relations could be very useful in this regard, and we plan to use these soft theorems to construct these shapes directly. Clearly the leading order soft theorems will not be sufficient as these don't have the power to constrain the full shapes, so rather one would need to use a collection of sub-leading soft theorems. Such relations will also require knowledge of correlators that mix the graviton and the curvature perturbation $\zeta$ \cite{Hinterbichler:2013dpa}.
    \item Given the previous point, it would be very interesting to construct mixed correlators. Writing down the general EFToI action and doing the computation will not be very efficient so one would need to develop bootstrap tools to construct these correlators. It is not yet clear how to do this in complete generality. Indeed, we expect $\zeta$ correlators to violate the MLT since the corresponding self-interactions are not manifestly local. One can again use soft theorems \cite{BBBB}, but another option would be to find a generalisation of the MLT that applies directly to $\zeta$ correlators. Such a generalisation should be possible given that the time integrals one needs to compute are the same as those of a spectator scalar, but one needs to effectively deal with the inverse Laplacians that arise when integrating out the non-dynamical parts of the metric. This perhaps requires a better understanding of locality in the presence of dynamical gravity. 
    \item More ambitiously, one might expect that three-point functions can be constrained by demanding consistency of higher-point functions such as the trispectrum. This is a familiar technique for scattering amplitudes where cubic couplings, and the spectrum, can be constrained by demanding that four-point amplitudes have only simple poles and factorise consistently on such poles \cite{TASI,PSS,Benincasa:2007xk}. Thanks to efforts of recent years we now have a solid understanding of the analytic properties of four-point cosmological correlators and such consistency conditions could be used to constrain the three-point functions that contribute to four-point functions. For the EFToI one would need to impose that the spectrum contains a single scalar and a massless graviton as the dynamical modes, and the Cosmological Optical Theorem \cite{COT} could provide a useful tool to yield constraints on three-point functions.  
\end{itemize}

\paragraph{Acknowledgements} We thank James Bonifacio, Harry Goodhew, Sadra Jazayeri, Austin Joyce, Enrico Pajer, Gui Pimentel and Dong-Gang Wang for helpful discussions. D.S. is supported by a UKRI Stephen Hawking Fellowship [grant number EP/W005441/1] and a Nottingham Research Fellowship from the University of Nottingham. G.C. acknowledges support from the Institute for Advanced Study. A.T. is supported by the Bell Burnell Graduate Scholarship Fund and the Cavendish (University of Cambridge). J.S. acknowledges support from STFC. For the purpose of open access, the authors have applied a CC BY public copyright licence to any Author Accepted Manuscript version arising.

\bibliographystyle{JHEP}
\bibliography{refsEFToI}

\end{document}